\documentclass[a4paper]{article}

\usepackage[margin=2.9cm]{geometry}
\usepackage[lofdepth,lotdepth]{subfig}
\usepackage{graphicx}
\usepackage{algorithm}
\usepackage{algpseudocode}
\usepackage{cite}
\usepackage{amssymb}
\usepackage{gensymb}
\usepackage{hyperref}

\usepackage[latin1]{inputenc}
\usepackage{tikz}
\usetikzlibrary{shapes,arrows,positioning,calc}

\newcommand{\dif}{{\rm d}}

\newcommand{\vJ}{{\bf J}}

\newcommand{\vE}{{\bf E}}
\newcommand{\vB}{{\bf B}}
\newcommand{\vA}{{\bf A}}
\newcommand{\vH}{{\bf H}}
\newcommand{\vT}{{\bf T}}

\newcommand{\rotDT}{\nabla\times\Delta{\bf T}}
\newcommand{\rotDTp}{\nabla' \times\Delta{\bf T}'}
\newcommand{\vr}{{\bf r}}

\newcommand{\half}{\frac{1}{2}}

\tikzstyle{startstop} = [rectangle, rounded corners, minimum width=3cm, minimum height=1cm,text centered, draw=black, fill=red!30]
\tikzstyle{io} = [trapezium, trapezium left angle=70, trapezium right angle=110, minimum width=3cm, minimum height=1cm, text centered, draw=black, fill=blue!30]
\tikzstyle{process} = [rectangle, minimum width=3cm, minimum height=1cm, align=left, draw=black, fill=orange!30]
\tikzstyle{empty} = [rectangle, align=left]
\tikzstyle{decision} = [diamond, minimum width=3cm, minimum height=1cm, align=center, draw=black, fill=green!30]
\tikzstyle{arrow} = [thick,->,>=stealth]

\begin{document}
%
\title{\flushleft \bf 3D magnetization currents, magnetization loop, and saturation field in superconducting rectangular prisms\footnote{This is an author-created, un-copyedited version of an article accepted for publication/published in Superconductor Science and Technology. IOP Publishing Ltd is not responsible for any errors or omissions in this version of the manuscript or any version derived from it. The Version of Record is available online at \url{https://doi.org/10.1088/1361-6668/aa69ed}.}}
\markboth{}{}

\author{E. Pardo\footnote{Corresponding author. Email: enric.pardo@savba.sk}, M. Kapolka\\
\normalsize{Institute of Electrical Engineering, Slovak Academy of Sciences,}\\
\normalsize{Dubravska 9, 84104 Bratislava, Slovakia}
}%

\date{\today}

\maketitle

\begin{abstract}
Bulk superconductors are used in both many applications and material characterization experiments, being the bulk shape of rectangular prism very frequent. However the magnetization currents are still mostly unknown for this kind of three dimensional (3D) shape, specially below the saturation magnetic field. Knowledge of the magnetization currents in this kind of samples is needed to interpret the measurements and the development of bulk materials for applications. This article presents a systematic analysis of the magnetization currents in prisms of square base and several thicknesses. We make this study by numerical modeling using a variational principle that enables high number of degrees of freedom. We also compute the magnetization loops and the saturation magnetic field, using a definition that is more relevant for thin prisms than previous ones. The article presents a practical analytical fit for any aspect ratio. For applied fields below the saturation field, the current paths are not rectangular, presenting 3D bending. The thickness-average results are consistent with previous modeling and measurements for thin films. The 3D bending of the current lines indicates that there could be flux cutting effects in rectangular prisms. The component of the critical current density in the applied field direction may play a role, being the magnetization currents in a bulk and a stack of tapes not identical.
\end{abstract}



\section{Introduction}
\label{s.intro}

Bulk superconductors are used in many applications, such as compact magnets for magnetic separation, superconducting motors and generators, and levitation systems \cite{hull04PIE,murakami07ACT,zhouD12SST,werfel12SST}. In addition, bulk samples are commonly used to study material properties of superconductors \cite{hecher15SST,mishev15SST}. The shape of rectangular prisms is frequently chosen for both applications and material characterization.

REBCO\footnote{REBCO stands for $RE$Ba$_2$Cu$_3$O$_{7-x}$, where $RE$ is a rare earth, typically Y, Gd or Sm.} superconducting bulks have been shown to trap large magnetic fields \cite{tomita03Nat,durrell14SST}, reaching 17.6 T \cite{durrell14SST}. Magnesium diboride (MgB$_2$) bulks are also promising, although the maximum trapped field is only 5.4 T \cite{fuchs13SST}. The advantages of MgB$_2$ are homogeneity, isotropy and, more important, higher power-law exponent in the relation between the electric field, $\vE$ and the current density $\vJ$. This is crucial for nuclear magnetic resonance (NMR) magnets, where the generated magnetic field should be highly stable.

Superconducting bulks can be magnetized in several ways. The most common is field-cool magnetization, consisting on placing a bulk in a DC magnet where the sample is above the critical temperature, $T_c$, decrease the temperature below $T_c$, and reduce the applied magnetic field from a certain initial value to zero. The maximum trapped field in the SC is reached when the initial applied field is above the saturation field \cite{ainslie15SST}. Alternatively, the sample could be cooled down at zero applied field, increase the applied field, and come back to zero. In this case, achieving the maximum trapped field requires a peak applied field that is twice the bulk-saturation field. These DC magnetization methods are often used for material characterization or testing the maximum possible trapped field. The present article provides an insight of this kind of magnetization. In certain applications, such as motors, it could be more convenient to use pulse magnetization \cite{ainslie15SST}. This method usually implies a combination of electro-magnetic and thermal processes, which is not the goal of this paper.


Although there are several works on 3D modelling of superconducting bulks \cite{acreview,ainslie15SST}, the magnetization currents in rectangular prisms of finite thickness remains mostly unknown. Infinite rectangular prisms in the CSM were analytically solved in \cite{chen89JAP}. Thin rectangular films have been studied in \cite{brandt95PRL,brandt95PRBa} and \cite{prigozhin98JCP,navau08JAP} for an isotropic power-law $\vE(\vJ)$ relation and the CSM, respectively. Computations for a rectangular prism with a hole have been published in \cite{pecher04ICS} for a power-law $\vE(\vJ)$ relation. Reference \cite{badia05APL} presented approximated solutions for a cube in the CSM, assuming square current paths. The trapped field of an array of rectangular prisms is computed in \cite{zhangM12SSTa}. 3D modelling has been applied to finite cylinders with holes \cite{lousberg09SST}; under transverse applied field, both as stand-alone superconductor \cite{campbell14SSTa} and interacting with ferromagnetic disks and rings \cite{fagnard16SST}; and under axial applied field but non-homogeneous critical current density \cite{komi09PhC,ainslie14SST}. Another 3D studied case is that of two rectangular superconducting filaments coupled by a normal metal in between \cite{grilli05IES}, although a single prism was not studied in that work.

Therefore, a systematic study of the magnetization process in finite rectangular prisms is needed. This article presents a detailed analysis of the 3D current flow, as well as the magnetization loops and saturation magnetic field. We focus on rectangular prisms of square base and several thicknesses. The analysis is done by numerical modeling using the Minimum Electro-Magnetic Entropy Production method in 3D (MEMEP3D), since it has been shown to be able to manage the high number of degrees of freedom necessary for this study \cite{memep3D}. 

Section \ref{s.mod} outlines the main features of the numerical model and the assumed material properties. We discuss the results in section \ref{s.results}, emphasizing the qualitative shape of the current paths below saturation (section \ref{s.J}). Section \ref{s.BpM} presents the magnetization loops and the saturation magnetic field calculated from them. There, we introduce a practical analytical fit of the saturation field that is useful for any aspect ratio, including thin prisms. We present our conclusions in section \ref{s.concl}.


\section{Model}
\label{s.mod}


\subsection{Material properties and physical situation}
\label{s.phmod}

In this work we consider an isotropic power law as
\begin{equation}
\vE(\vJ)=E_c \left ( \frac{|\vJ|}{J_c} \right )^n\frac{\vJ}{|\vJ|},
\label{EJ}
\end{equation}
where $E_c$ is an arbitrary constant, usually $10^{-4}$ V/m, $J_c$ is the critical current density, and $n$ is the power-law exponent. The limit of $n\to\infty$ corresponds to the isotropic critical-state model (CSM), which assumes a multi-valued $\vE(\vJ)$ relation.

In this article we assume uniform applied fields\footnote{We do not take magnetic materials into account, and hence the magnetic field and magnetic flux density are proportional $\vH=\vB/\mu_0$ being $\mu_0$ the void permeability. In the text, we use ``magnetic field" to refer to both the magnetic field and magnetic flux density.}, $\vB_a$; although the presented variational principle is also valid for non-uniform applied fields. We consider that the applied field follows the $z$ direction and is generated by a long racetrack coil in the $y$ direction and high in the $z$ direction. The resulting applied vector potential $\vA_a$ in Coulomb's gauge, defined as Appendix B in \cite{acreview}, is
\begin{equation}
\vA_a(\vr)\approx B_ax{\bf e}_y,
\label{Aalong}
\end{equation}
where $B_a$ is such that $\vB_a=B_a{\bf e}_z$, and ${\bf e}_y,{\bf e}_z$ are the unit vectors in the $y$ and $z$ directions, respectively.


\subsection{Numerical method}
\label{s.varprin}	

We use the MEMEP3D variational method, detailed in \cite{memep3D}. This method is valid for any combination of applied field and transport current and any vector $\vE(\vJ)$ relation, including the case of magnetic-field dependent parameters in the $\vE(\vJ)$ relation like $J_c$ and $n$.

Outlining, for no transport current, the current density is taken as a function of the effective magnetization $\vT$ as
\begin{equation}
\vJ=\nabla\times\vT.
\end{equation}
When increasing the time from a certain $t_0$ to $t_0+\Delta t$, the applied vector potential changes by $\Delta \vA_a$, which causes a change in $\vT$ by $\Delta\vT$. Then, $\vJ=\vJ_0+\rotDT$, where $\vJ_0$ is $\vJ$ at time $t_0$. This $\Delta\vT$ is obtained by numerically minimizing the functional
\begin{eqnarray}
L[\Delta\vT] & = & \int_V\dif V\int_V\dif V' \frac{\mu_0}{8\pi\Delta t} \frac{(\rotDT)\cdot(\rotDTp)}{|\vr-\vr'|} \nonumber\\
& + & \int_V\dif V \Bigg( \frac{\Delta\vA_a}{\Delta t}\cdot\rotDT + U(\vJ_0+\rotDT) \Bigg), 
\label{LTint}
\end{eqnarray}
where the dissipation factor $U(\vJ)$ is defined as
\begin{equation}
U(\vJ)\equiv\int_0^{\bf J}\dif \vJ'\cdot\vE(\vJ').
\label{U}
\end{equation}
For the power-law $\vE(\vJ)$ relation of (\ref{EJ}), the dissipation factor becomes
\begin{equation}
U(\vJ)=\frac{E_cJ_c}{n+1}\left ( \frac{|\vJ|}{J_c} \right )^{n+1}
.
\end{equation}

As demonstrated in \cite{memep3D}, this functional always presents a minimum and that minimum is unique. Minimizing this functional is equivalent to solving Faraday's law
\begin{equation}
\nabla\times\vE=-\frac{\partial\vB}{\partial t}.
\end{equation}
We minimize the functional of (\ref{LTint}) by the iterative parallel algorithm of \cite{memep3D} in a self-programmed implementation.

\subsection{Magnetization}
\label{s.mag}

The average magnetization ${\bf M}$ is defined as the total magnetization $\bf m$ per unit sample volume $V$ as ${\bf M}={\bf m}/V$. The magnetic moment is
\begin{eqnarray}
{\bf m  } & = & \half\int\dif V\ \vr\times\vJ. \label{mrJ}
\end{eqnarray}
We evaluate the integral in (\ref{mrJ}) by assuming $\vr\times\vJ$ uniform in the cells and taking the values at the cells center, being $\vJ$ interpolated there.


\section{Results and discussion}
\label{s.results}

This section presents the current density, saturation field and magnetization loops for rectangular prisms with several height-to-width aspect ratios $c\equiv d/w$ and square base, being $w$ and $d$ the sample width and thickness. For all cases, we consider a power-law exponent of 100, unless stated otherwise.


\subsection{Current density}
\label{s.J}

\begin{figure}[ptb]
\centering
{\includegraphics[trim=0 30 0 0,clip,width=5.5 cm]{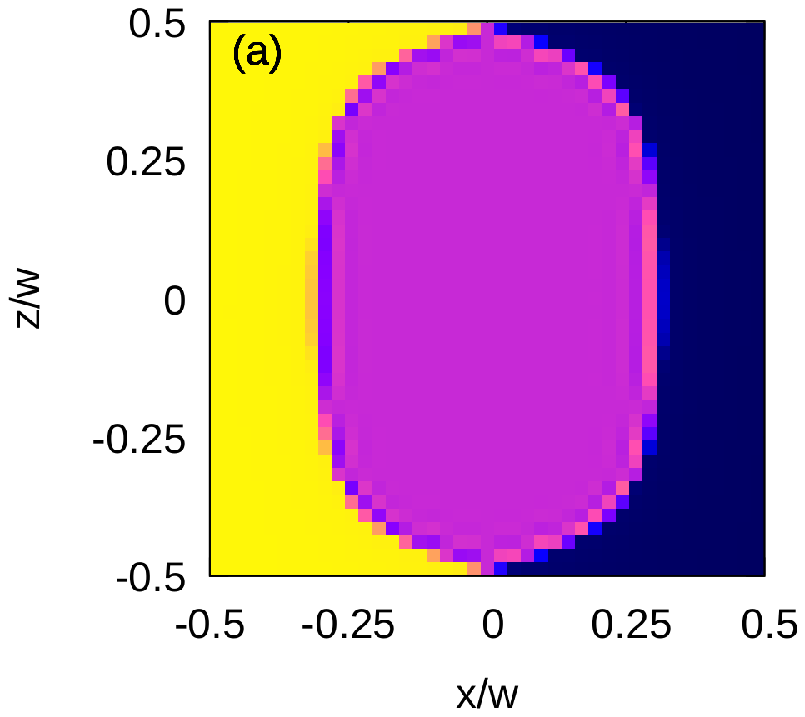}}\\
{\includegraphics[trim=0 43 0 0,clip,width=5.5 cm]{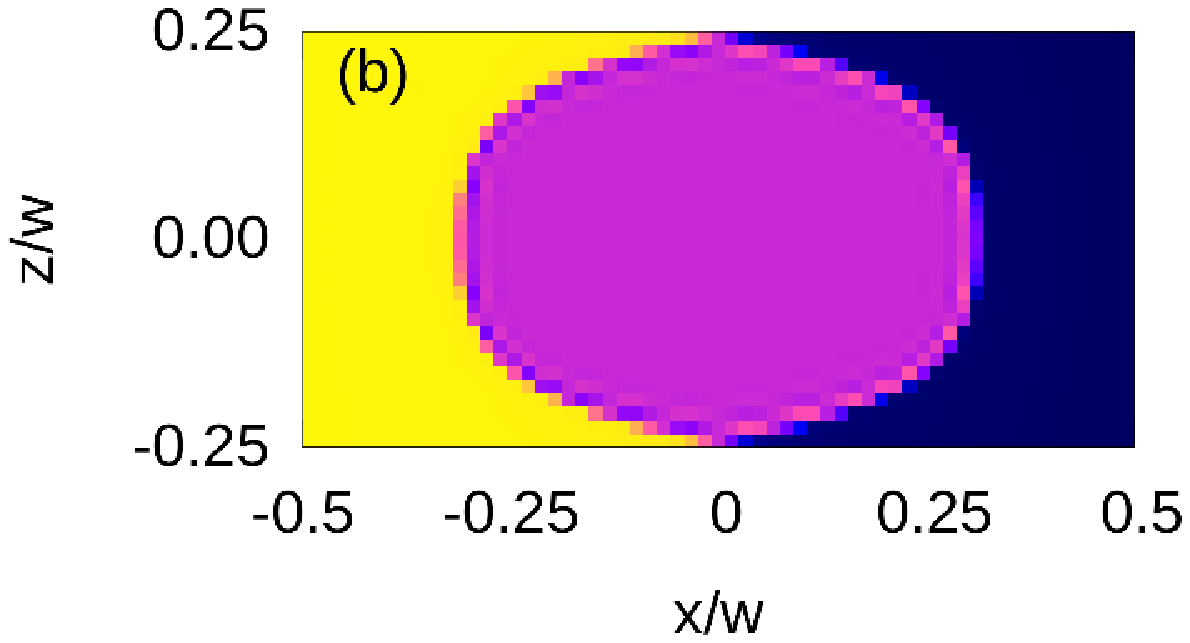}}\\
{\includegraphics[trim=0 43 0 0,clip,width=5.5 cm]{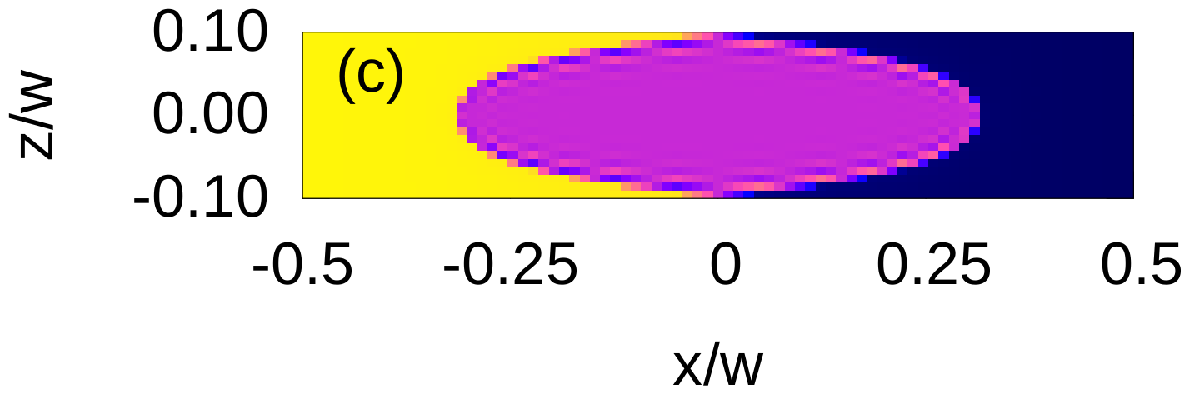}}\\
{\includegraphics[trim=0 -10 0 0,clip,width=5.5 cm]{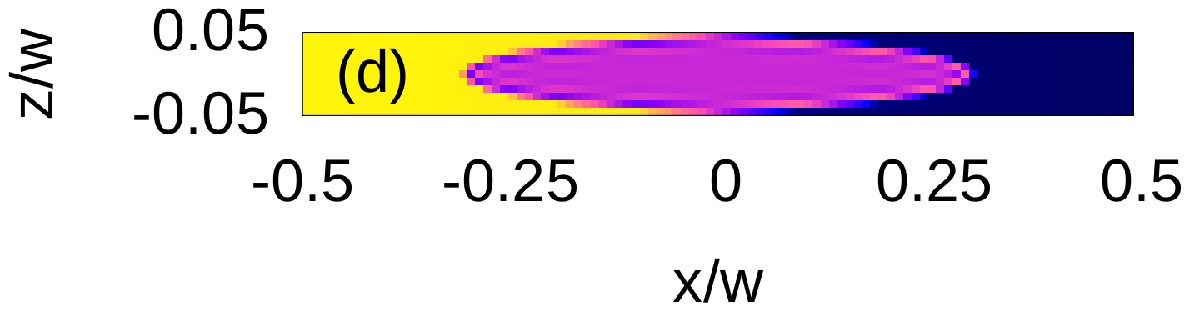}}\\
{\includegraphics[trim=0 0 0 58,clip,width=5.5 cm]{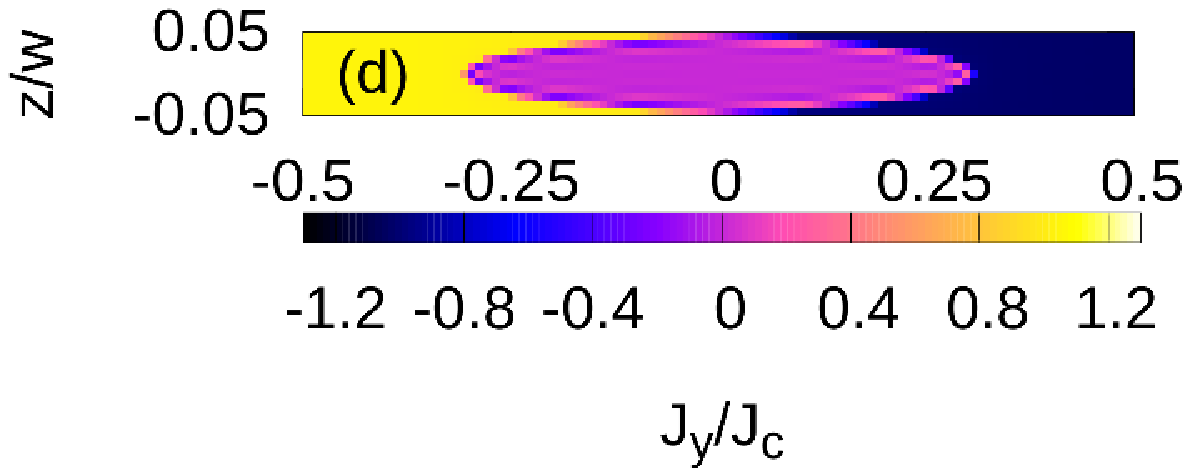}}
\caption{$J_y$ at the midplane ($y=0$) for rectangular prisms of square base and thickness-over-width ratio $c=d/w=$1,0.5,0.2,0.1 in (a,b,c,d), respectively. For all cases, the applied field is $B_a=0.484B_s$, being $B_s$ the saturation field in table \ref{t.Bp}.}\label{cubecJy.fig}
\end{figure}

\begin{figure}[ptb]
\centering
{\includegraphics[trim=0 26 0 0,clip,width=6.5 cm]{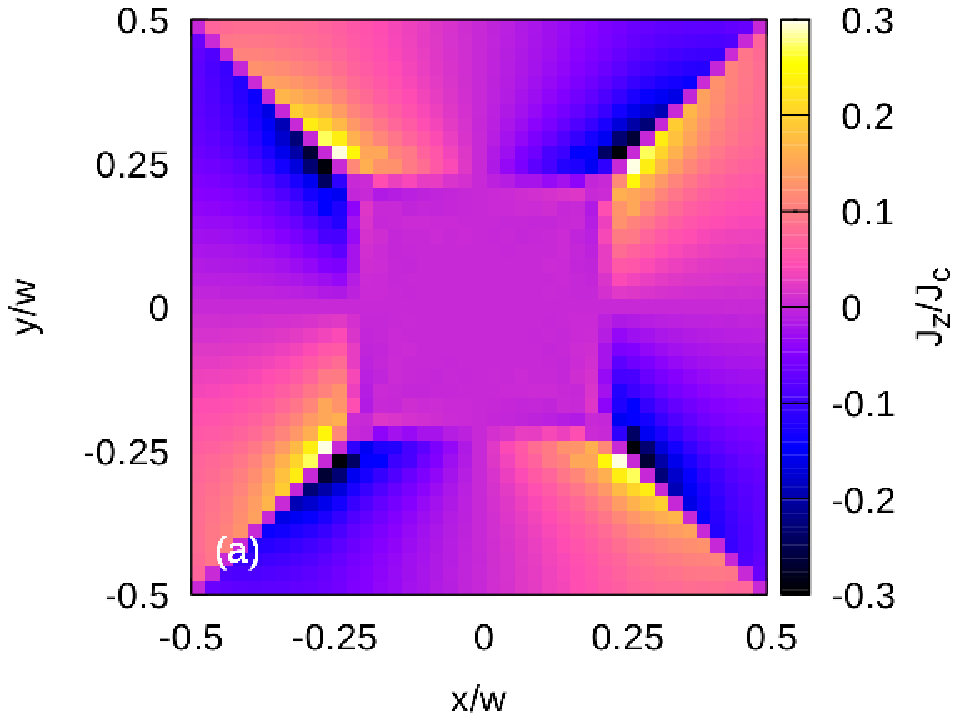}}\\
{\includegraphics[trim=0 26 0 0,clip,width=6.5 cm]{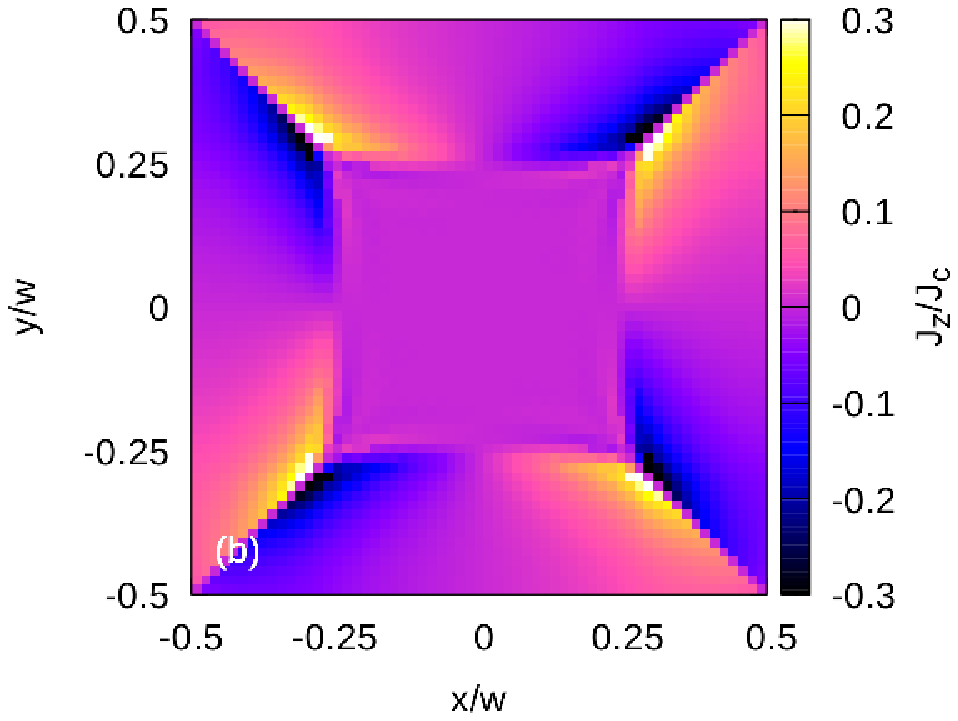}}\\
{\includegraphics[trim=0 26 0 0,clip,width=6.5 cm]{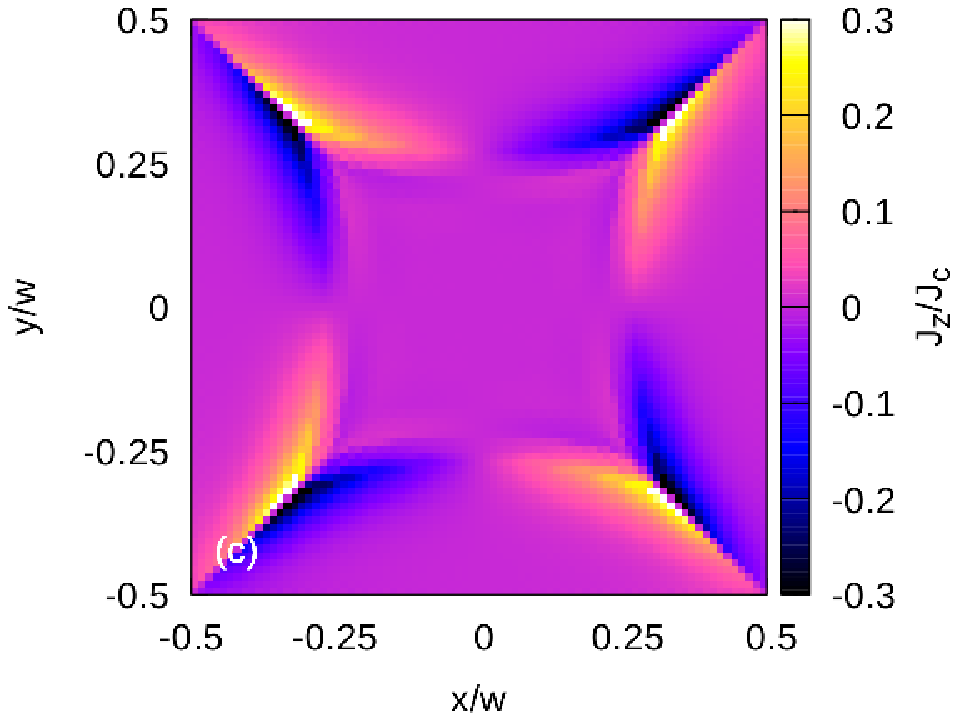}}\\
{\includegraphics[trim=0 0 0 0,clip,width=6.5 cm]{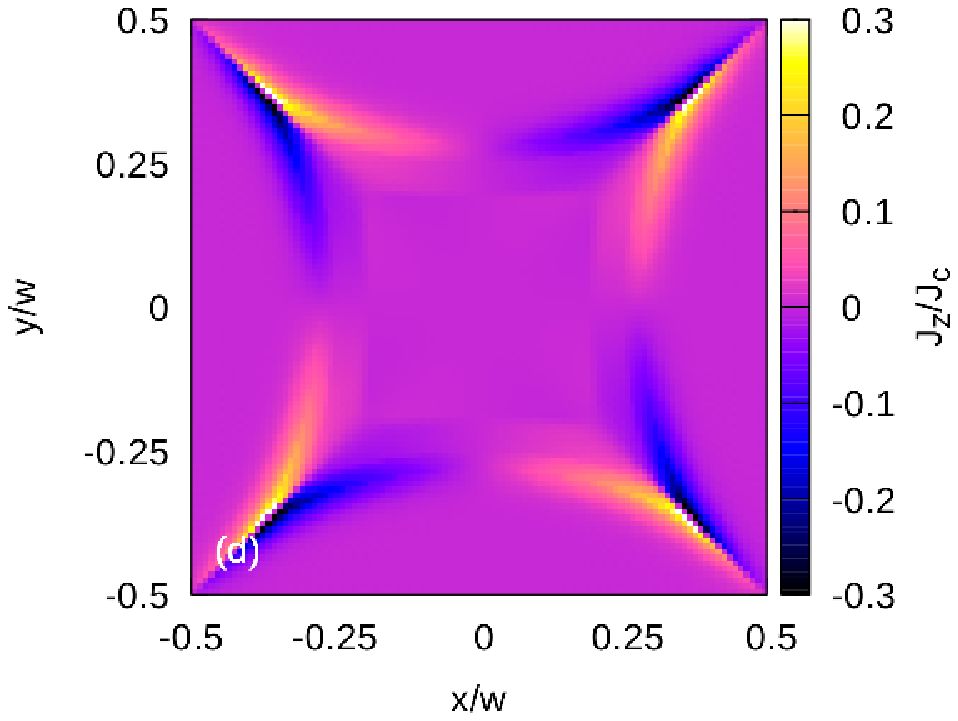}}
\caption{The maximum $J_z$ in a rectangular prism stays with a value around 30 \% $J_c$ for all aspect ratios, $c=d/w=$1,0.5,0.2,0.1 in (a,b,c,d), respectively. The maps, of constant $z$, are for the planes where $|J_z|$ is maximum, being at $z/d=$0.11,0.18,0.21,0.23 for $c=$1,0.5,0.2,0.1, respectively. The applied field is the same as in figure \ref{cubecJy.fig}.
}\label{cubecJz.fig}
\end{figure}

\begin{figure}[ptb]
\centering
{\includegraphics[trim=0 0 0 10,clip,height=4.8 cm]{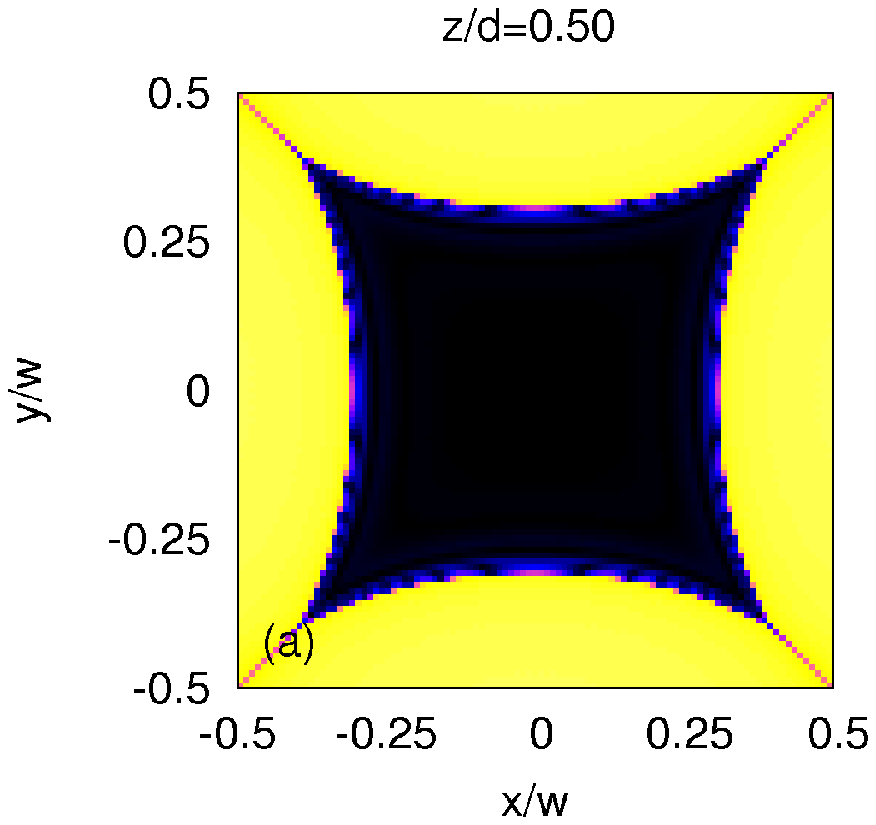}} 
{\includegraphics[trim=0 0 0 10,clip,height=4.8 cm]{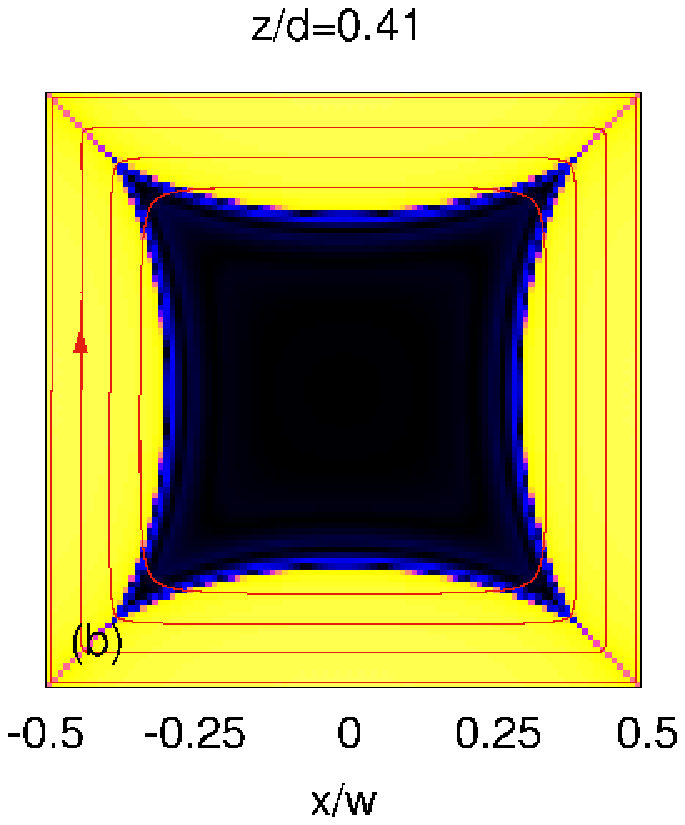}}
{\includegraphics[trim=0 0 0 10,clip,height=4.8 cm]{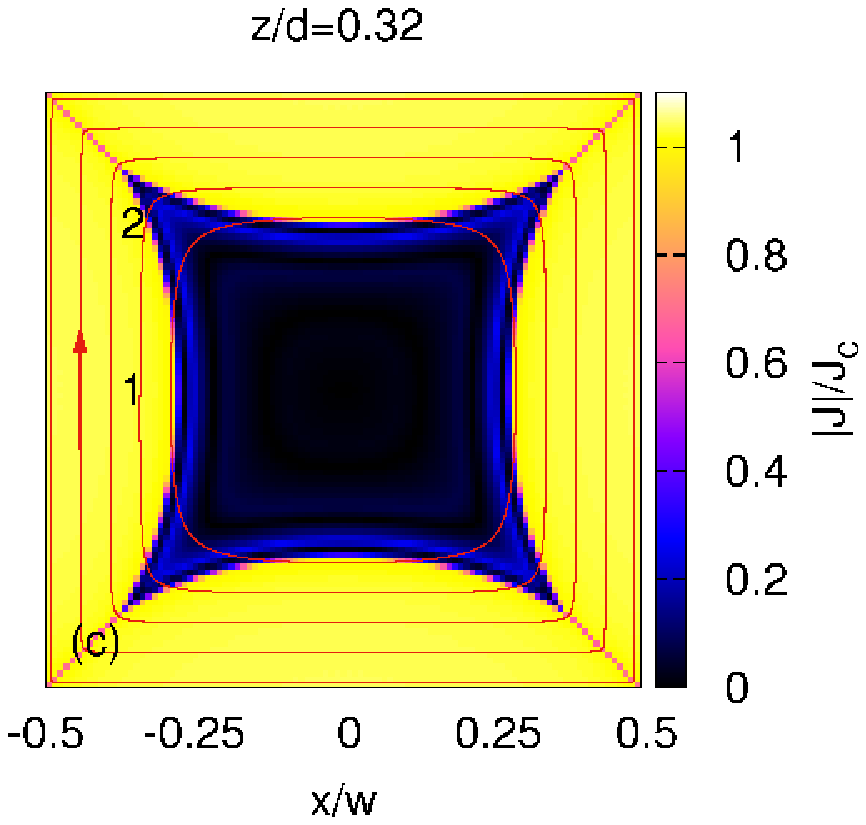}}\\ 
{\includegraphics[trim=0 0 0 10,clip,height=4.8 cm]{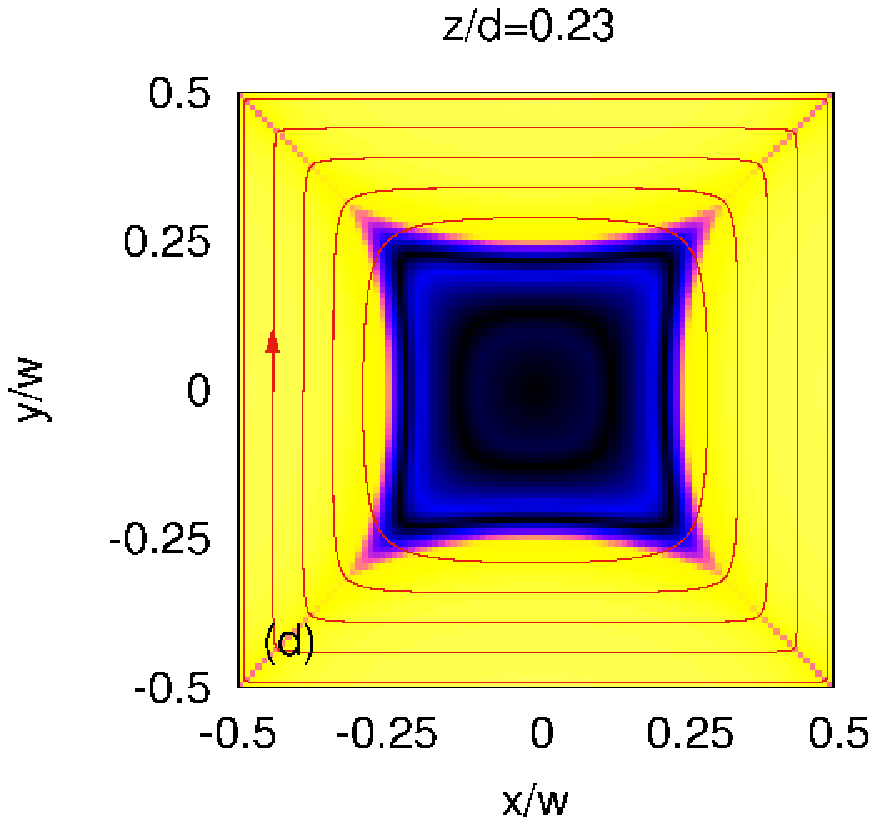}} 
{\includegraphics[trim=0 0 0 10,clip,height=4.8 cm]{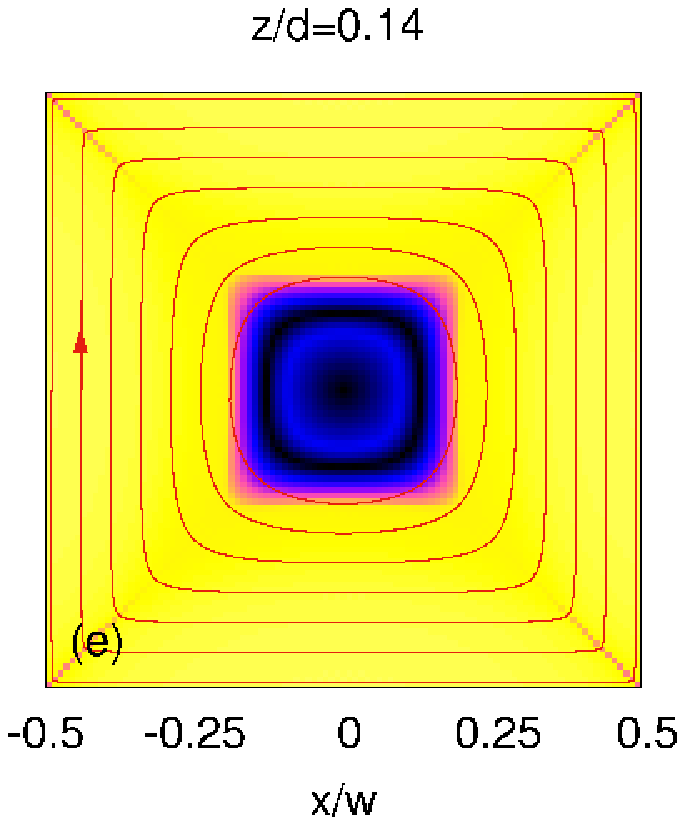}} 
{\includegraphics[trim=0 0 0 10,clip,height=4.8 cm]{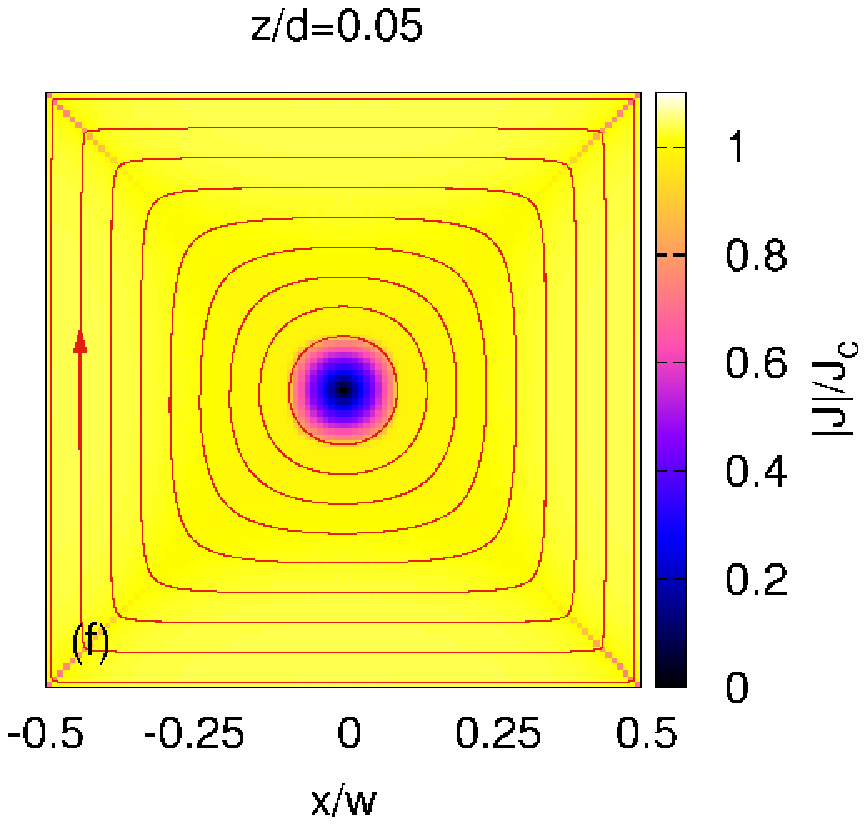}} 
\caption{The current flux lines in a thin prism ($d/w=$0.1) do not follow square loops [plots for $z/d=$0,-0.09,-0.18,-0.27,-0.36,-0.45 in (a,b,c,d,e,f), respectively]. The graphs show the projection of the 3D current lines in the $xy$ plane and the magnitude of the current density (colormap), since the 3D flux lines show the direction of the current density but not its magnitude. Current lines start at $y=0$ and the $z$ coordinate of the plotted map but bend in the $z$ direction, belonging to other $z$ layers close to the diagonals. Computed case for power-law exponent 100 and $B_a=0.484B_s$.} \label{cubec0.1.fig}
\end{figure}

\begin{figure}[ptb]
\centering
{\includegraphics[trim=0 0 0 10,clip,height=4.8 cm]{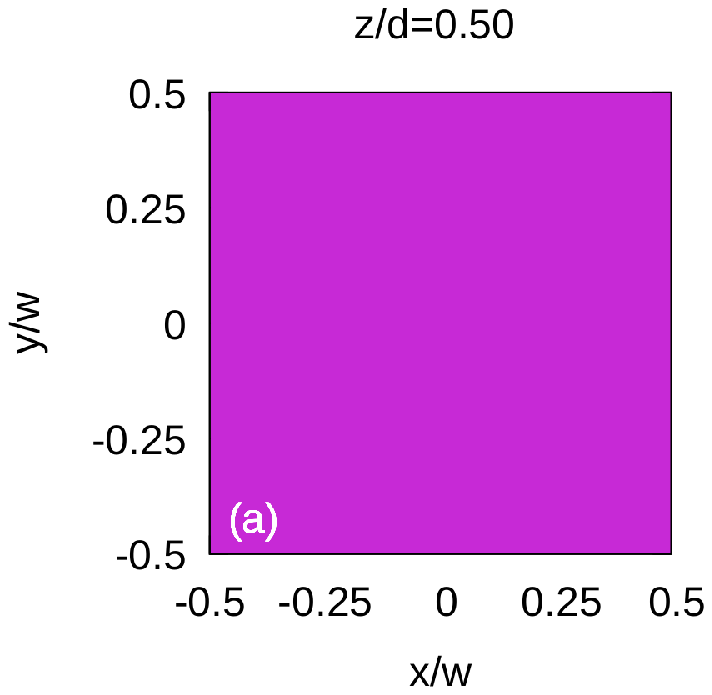}} 
{\includegraphics[trim=0 0 0 10,clip,height=4.8 cm]{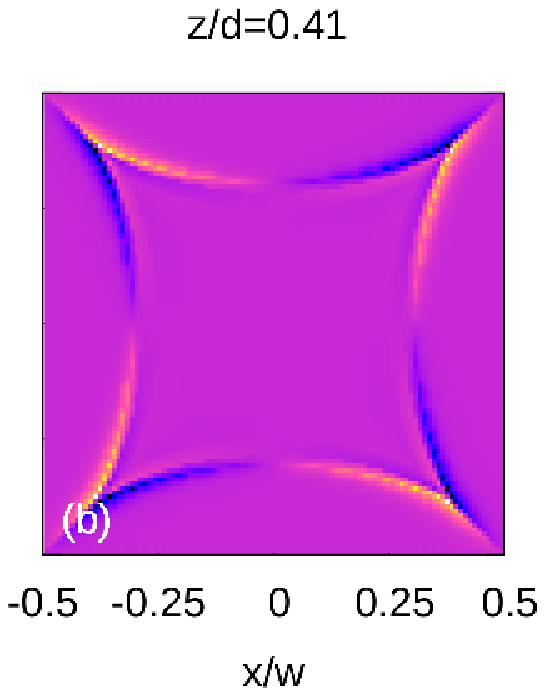}} 
{\includegraphics[trim=0 0 0 10,clip,height=4.8 cm]{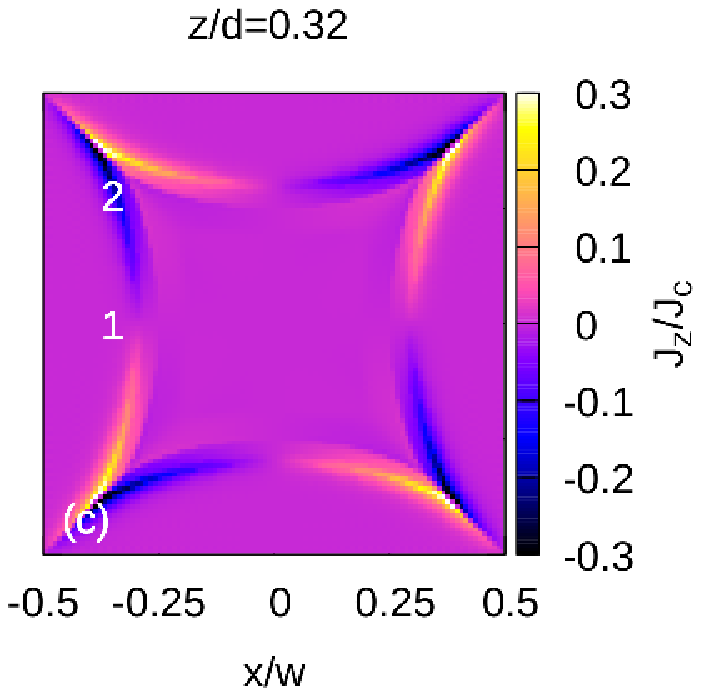}}\\
{\includegraphics[trim=0 0 0 10,clip,height=4.8 cm]{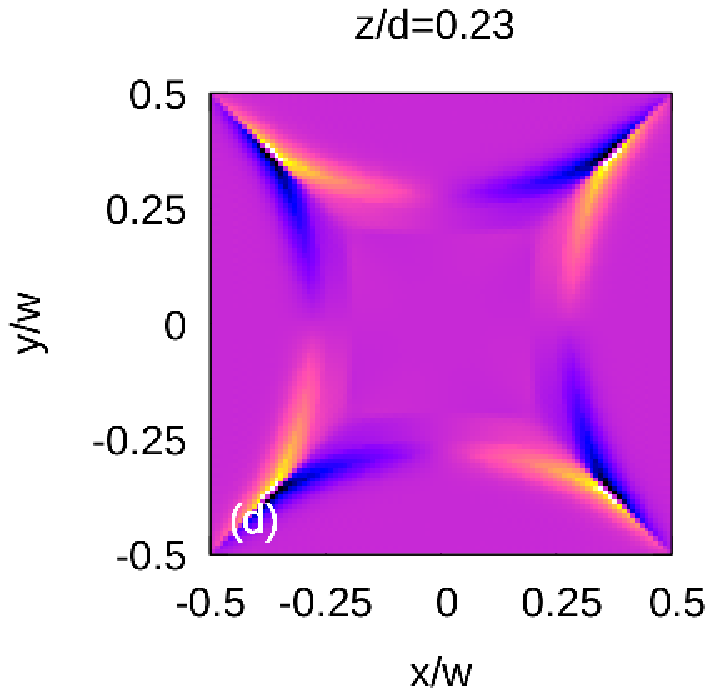}} 
{\includegraphics[trim=0 0 0 10,clip,height=4.8 cm]{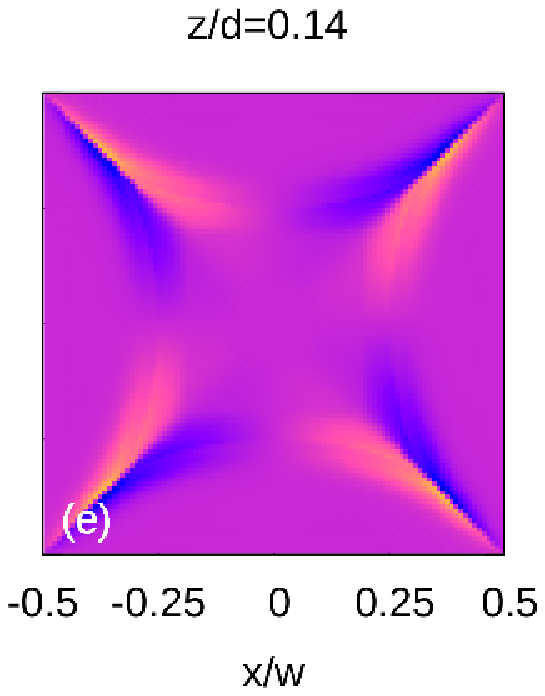}} 
{\includegraphics[trim=0 0 0 10,clip,height=4.8 cm]{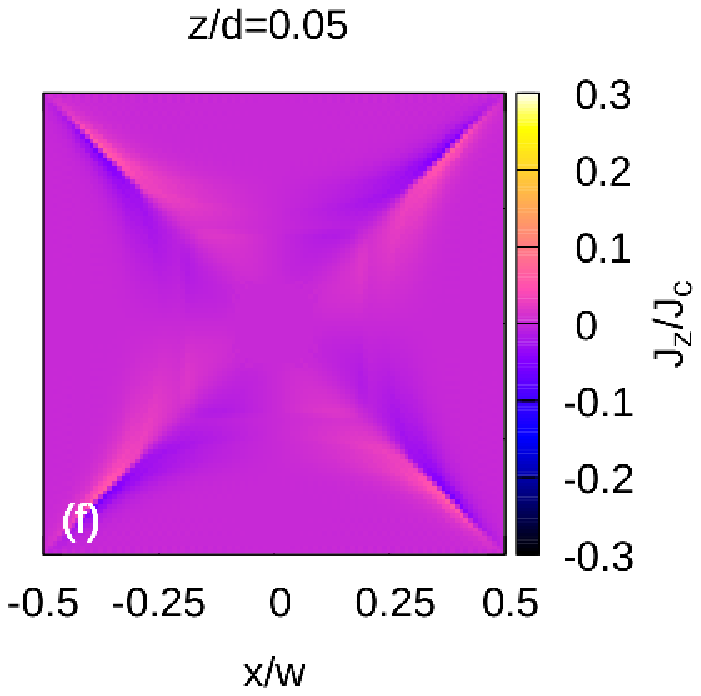}} 
\caption{In a thin prism ($d/w=0.1$), there appear regions with $J_z$ at all heights except a the midplane $z=0$ [plots for $z/d=$0,-0.09,-0.18,-0.27,-0.36,-0.45 in (a,b,c,d,e,f), respectively], causing current lines with 3D bending (see figure \ref{cubec0.1.fig}). Same computed situation as figure \ref{cubec0.1.fig}.}\label{cubec0.1Jz.fig}
\end{figure}

\begin{figure}[ptb]
\centering
{\includegraphics[trim=10 20 25 20,clip,width=10.0 cm]{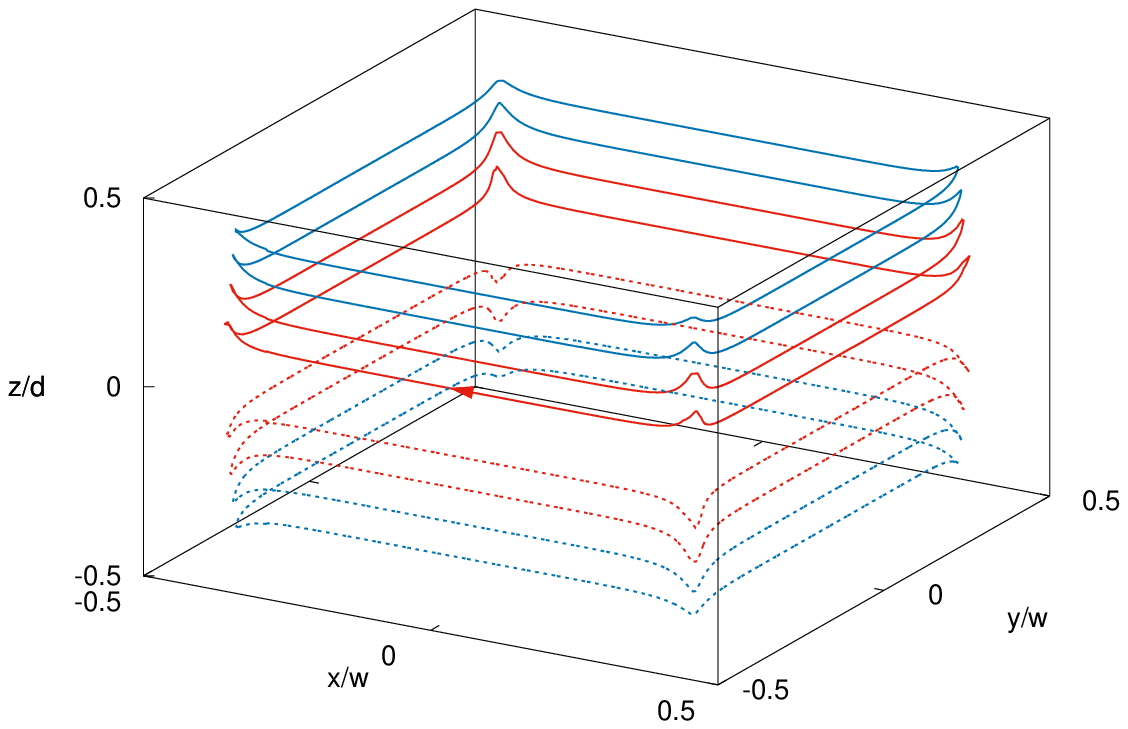}}
{\includegraphics[trim=10 20 25 20,clip,width=10.0 cm]{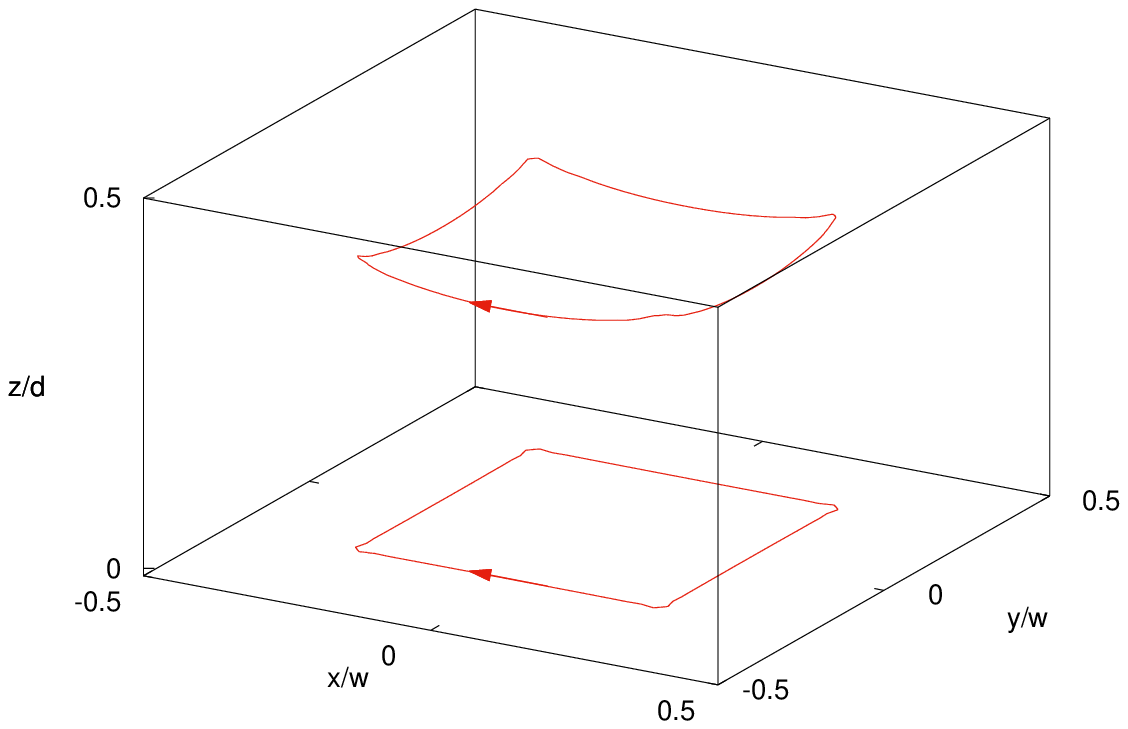}}
\caption{The current flux lines present 3D bending. Shown is a thin film with $d/w=0.1$ (top) and cube (bottom). Close to the center, the $z$ bending for a cube is negligible. Different colors and line types are used in the top graph for clarity.}
\label{Jline3d.fig}
\end{figure}

\begin{figure}[ptb]
\centering
{\includegraphics[trim=15 4 15 10,clip,height=4.8 cm]{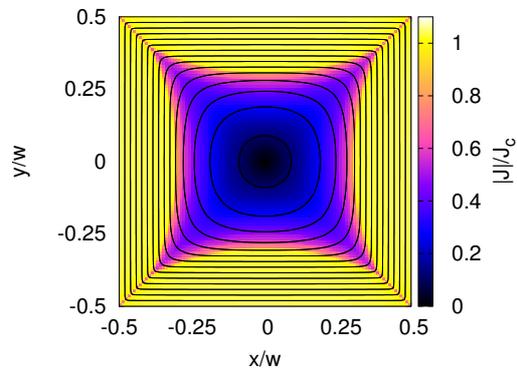}}
\caption{The thickness-average current density for the sample with aspect ratio 0.1 (same situation as figure \ref{cubec0.1.fig}) is consistent with both previous calculations for thin films \cite{brandt95PRBa,prigozhin98JCP} and current density extracted from magnetic field imaging \cite{jooss02RPP,romerosalazar10PRB,wells17ScR}. The lines are current lines for the thickness-averaged $\bf J$.}
\label{Jav.fig}
\end{figure}

\begin{figure}[ptb]
\centering
{\includegraphics[trim=0 -10 0 0,clip,height=4.6 cm]{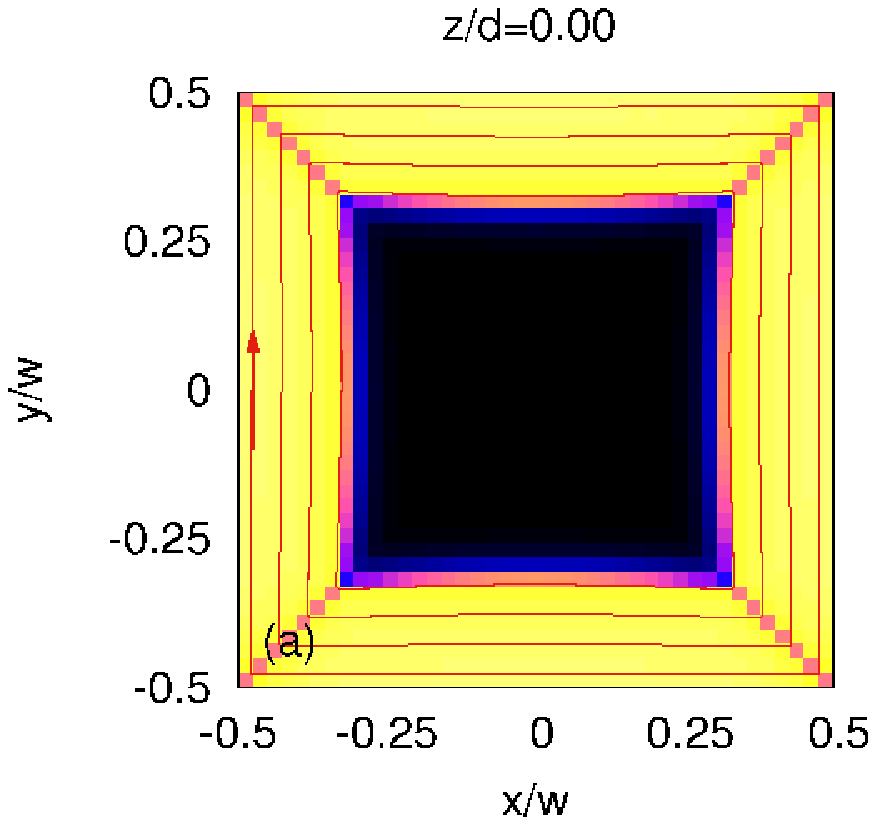}}
{\includegraphics[trim=0 -10 0 0,clip,height=4.6 cm]{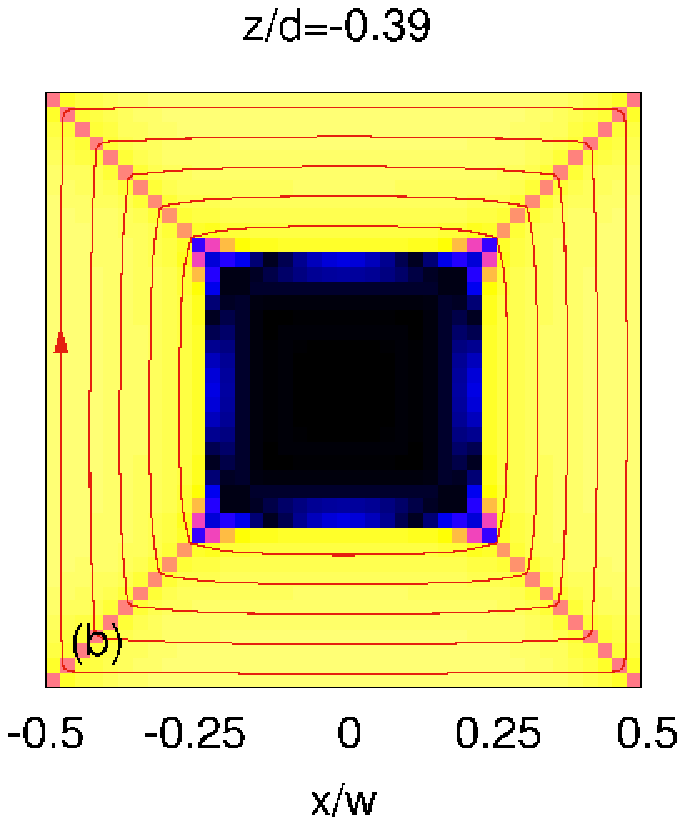}}  
{\includegraphics[trim=0 -10 0 0,clip,height=4.6 cm]{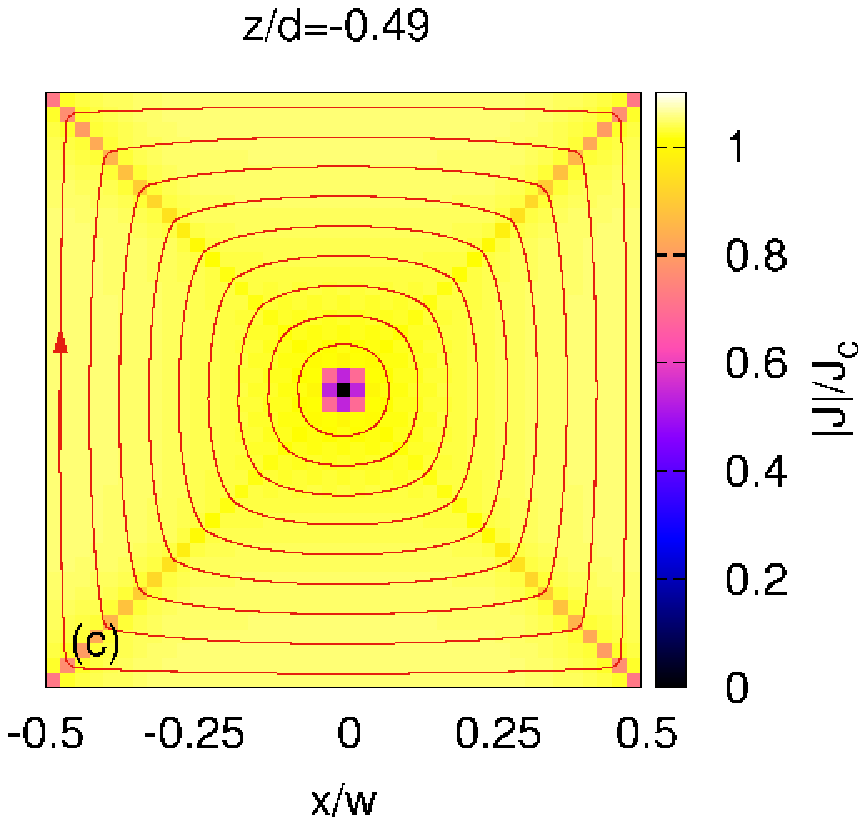}} \\
{\includegraphics[trim=0 0 0 0,clip,height=4.5 cm]{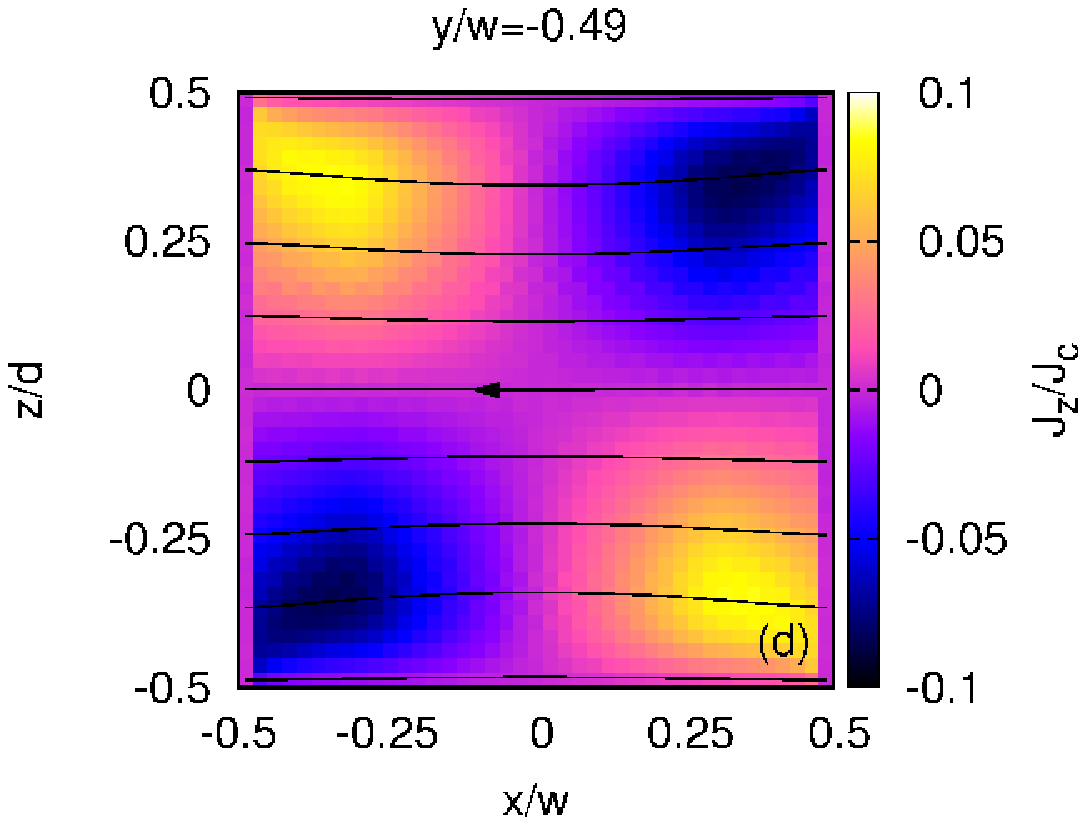}}
{\includegraphics[trim=0 -20 0 0,clip,width=5 cm]{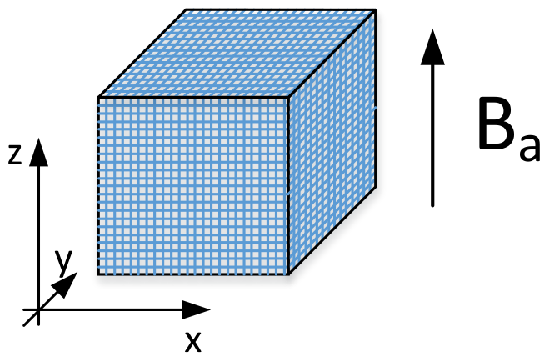}}
\caption{Current density magnitude (a,b,c) and $J_z$ (d) at several cross-sections of a cube, $d/w=1$. The lines are 3D current flux lines projected on the plotted plane that start at $y=0$ in (a,b,c) and $x=0.5w$ in (d), representing the direction of the current density but not its magnitude. Computed case for power-law exponent 100 and applied field $B_a=0.484 B_s$, being $B_s$ the saturation field given in table \ref{t.Bp}. The sketch shows the taken geometry and axis.}\label{cubeJc.fig}
\end{figure}

We compare the current density in prisms of several aspect ratios and the same applied field relative to the saturation field $B_a/B_s=0.484$. We define the $z$ axis as that of the applied field (see sketch in figure \ref{cubeJc.fig}). The saturation field, $B_s$ is presented in section \ref{s.BpM}.

The $J_y$ component of the current density at the central plane of constant $y$ for several aspect ratios is at figure \ref{cubecJy.fig}. The highest penetration depth is at the top and bottom, being the smallest at the middle. The roughly flat part of the penetration front close to the central $z$ decreases with the prism height.

Up to now, the current density is qualitatively similar to cylinders \cite{brandt98PRBa,sanchez01PRB}. Two main differences are the presence of a non-zero $J_z$ component (see figure \ref{cubecJz.fig}) and current paths that do not always follow the shape of the sample boundary, and hence they are non-square (figures \ref{cubec0.1.fig}, \ref{Jline3d.fig} and \ref{cubeJc.fig}c).

The $J_z$ component reaches values as high as 30 \% of ${J_{c}}$ (figure \ref{cubecJz.fig}). The highest magnitude of ${J_{z}}$ is close to the diagonal of the prisms. This $J_z$ bends the current flux lines, as seen in the 3D current loops in figure \ref{Jline3d.fig}. The cause of this $J_z$ component is the self-field. In cylinders, the radial component of the self-field, perpendicular to the current loops, is balanced by higher current penetration close to the ends \cite{sanchez01PRB}. That is possible thanks to the cylindrical symmetry, which causes that the radial field is uniform in any circular loop. This no longer applies to rectangular prisms. The magnetic field created by rectangular loops at the diagonal is higher than closer to the straight parts at the same distance from the lateral faces \cite{brandt95PRL}. Thus, higher current penetration close to the ends following rectangular loops cannot fully cancel the self-field. Close to the diagonals, the additional perpendicular self-field pointing inwards is canceled by a $J_z$ component that changes its sign at the diagonal. For applied fields well above the saturation field, the self-field is not relevant, and hence the current paths follow rectangular loops in the whole sample.

Lets analyze in detail the thinnest sample, with $d/w=0.1$. Figures \ref{cubec0.1.fig} and \ref{cubec0.1Jz.fig} show the maps of $|\vJ|$ and $J_z$ for all computed heights in the lower half of the prism. Figure \ref{cubec0.1.fig} also shows the projections in the $xy$ plane of the 3D current flux lines. An interesting issue is that the current loops do not close within the same plane. The roughly straight segments 	close to one side (point 1 at figures \ref{cubec0.1.fig}c and \ref{cubec0.1Jz.fig}c) bend downwards when approaching to the diagonal (point 2 at figures \ref{cubec0.1.fig}c and \ref{cubec0.1Jz.fig}c), where $J_z<0$, reach a lower layer, where $|J|$ is slightly below $J_c$, and bend in the $xy$ plane, describing roughly a circular arc. Thus, the current follows non-square loops, while keeping a star-shaped current-free zone at the planes close to the center. The $z$ bending of the current lines is compatible with $J_z=0$ exactly at the mid-plane because the $z$-bending for the current lines at slightly above and slightly below $z=0$ is in opposite directions (top graph of figure \ref{Jline3d.fig}). Current loops close to the prism side surfaces are still nearly square (figure 3).

Non-square current paths in thin films with regions with $|J|<J_c$ were found in \cite{brandt95PRBa,prigozhin98JCP} from numerical modeling and in \cite{jooss02RPP,romerosalazar10PRB,wells17ScR} from inversion of measured magnetic field maps. Although thin film models assume uniform $\vJ$ along the film thickness, their predictions are valid for the thickness-averaged $\vJ$ of samples with small but finite thickness. Our calculations for the thinnest sample ($d/w=0.1$) qualitatively agree with the thin-film calculations and measurements (figure \ref{Jav.fig}). The thickness-averaged current paths from thin films cannot be obtained by superposing square paths at several heights because they always result in square current paths. Therefore, non-square current paths need to exist in prisms, at least below their saturation field. In addition, the arguments above justifying the non-zero $J_c$ are valid for any aspect ratio. This constrasts with earlier predictions in \cite{badia05APL} for a cube, where in-plane square loops were assumed. This assumption was supported by taking into account that $|J|$ follows $|J|=J_c$ or 0 only, while the CSM allows any $|J|\le J_c$. Current densities with magnitude slightly below $J_c$ are enough to bend $\vJ$ vertically and obtain the necessary $J_z$ to shield the self-field. Assuming rectangular current loops should still produce fairly accurate results of the magnetic moment, although the obtained magnetic field at the surface will present significant errors. 

\begin{figure}[ptb]
\centering
{\includegraphics[trim=0 -10 0 0,clip,height=4.6 cm]{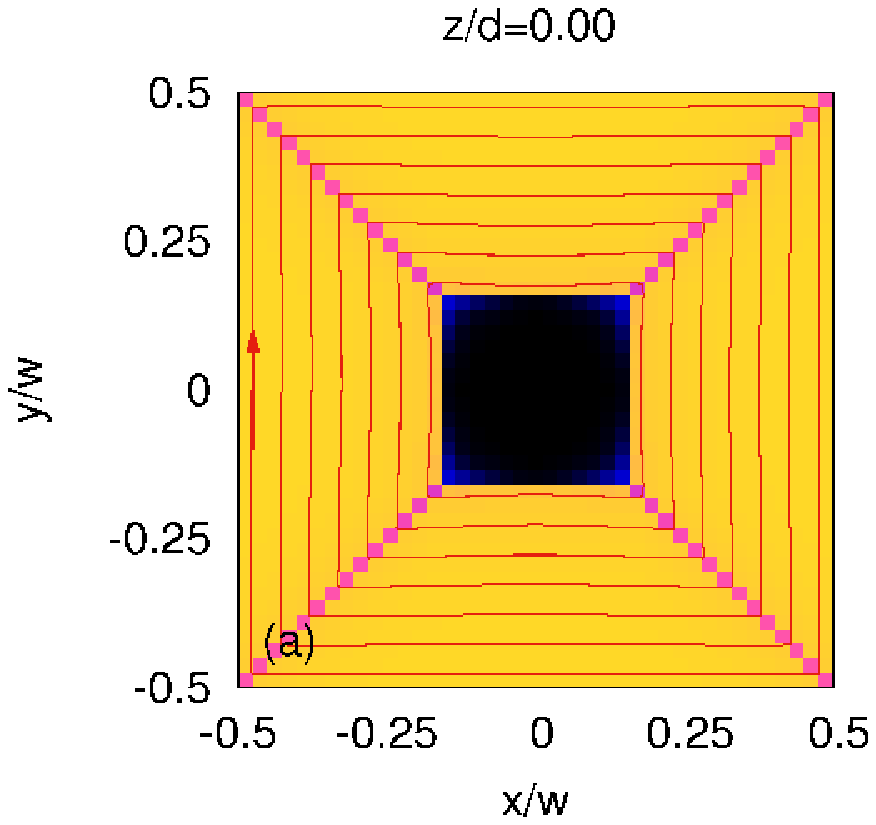}}
{\includegraphics[trim=0 -10 0 0,clip,height=4.6 cm]{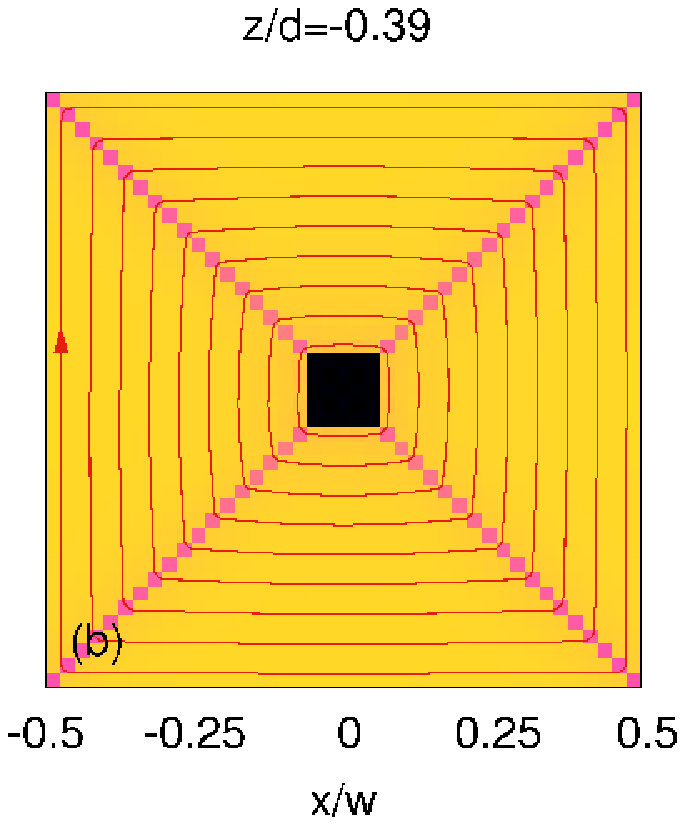}}  
{\includegraphics[trim=0 -10 0 0,clip,height=4.6 cm]{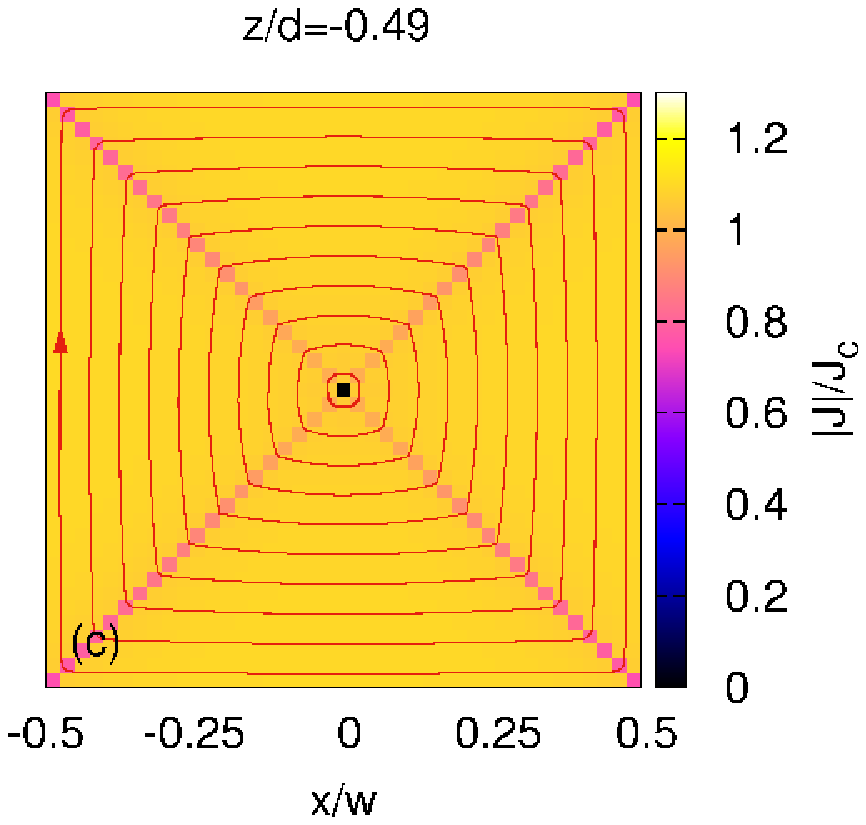}} \\
{\includegraphics[trim=0 0 0 0,clip,height=4.5 cm]{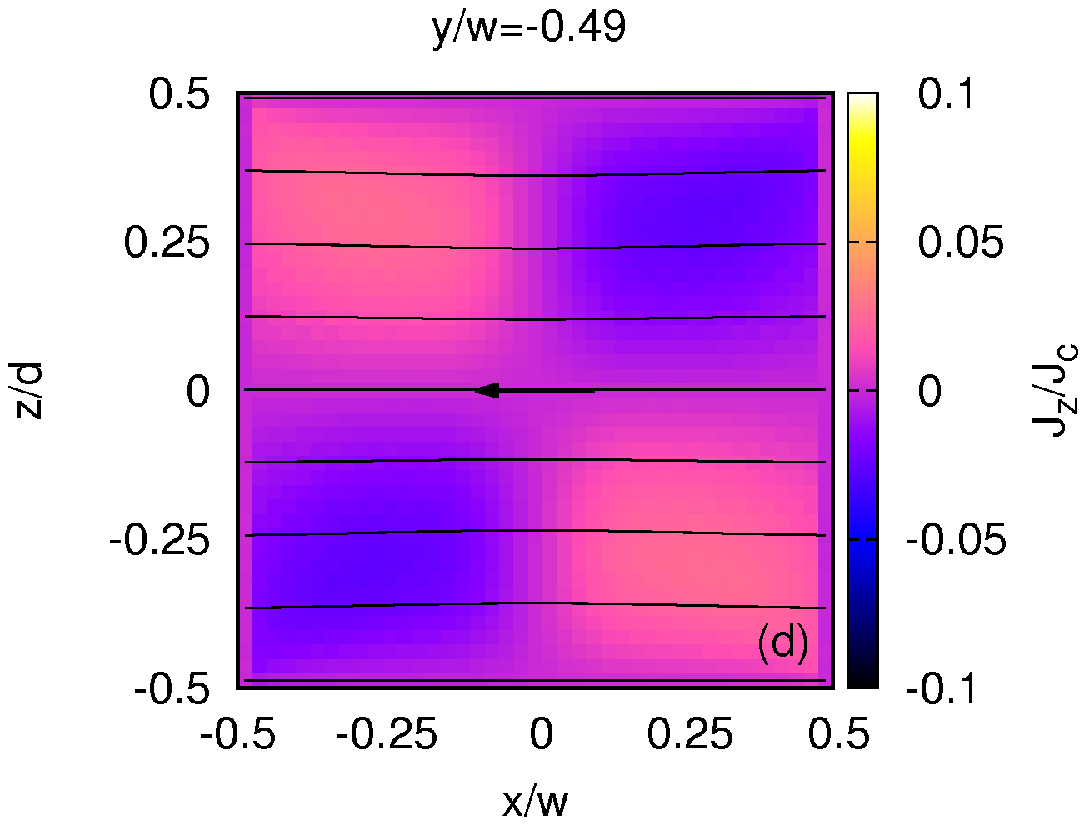}} 
{\includegraphics[trim=0 0 0 0,clip,height=4.5 cm]{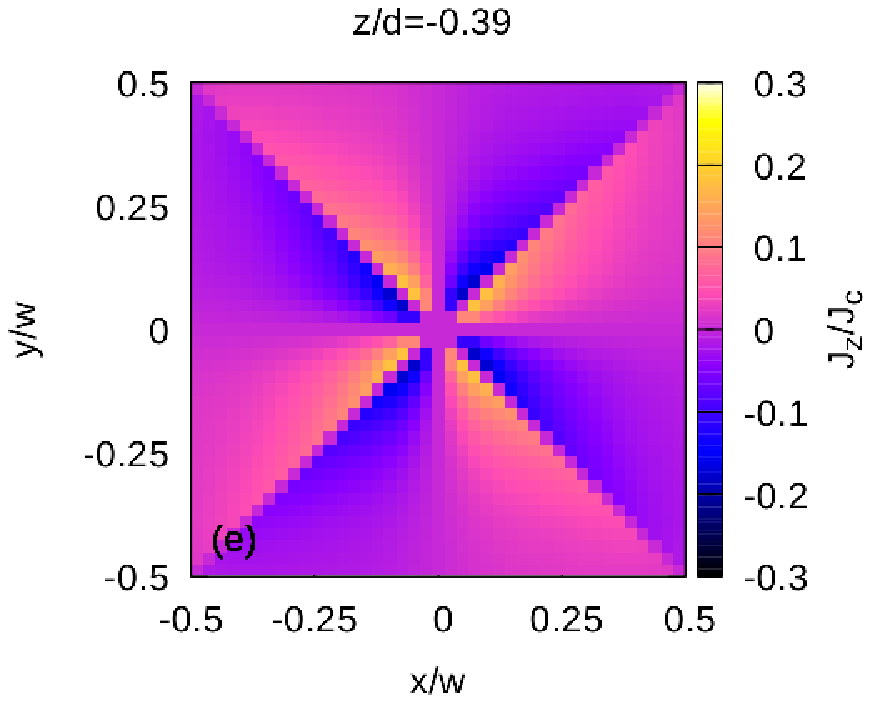}}
\caption{The same as figure \ref{cubeJc.fig} for (a,b,c,d) and figure \ref{cubecJz.fig}a for (e) but for an applied field $B_a=B_s$; defined as $M(B_a)=0.99M_s$, being $M_s$ the saturation magnetization.}\label{cubeJcBp.fig}
\end{figure}

\begin{figure}[ptb]
\centering
{\includegraphics[trim=0 0 0 0,clip,height=4.8 cm]{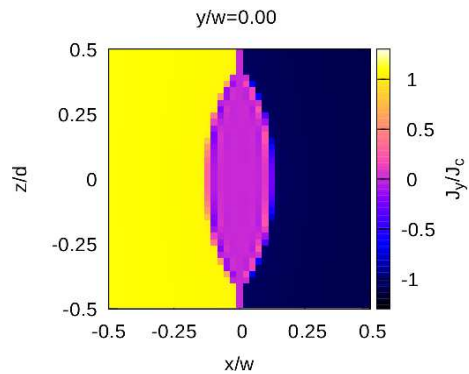}} 
\caption{The same as figure \ref{cubecJy.fig}a but for for $B_a=B_s$.}\label{cubeJcBpJzJy.fig}
\end{figure}

\begin{figure}[tbp]
\centering
{\includegraphics[trim=0 -10 0 0,clip,height=4.6 cm]{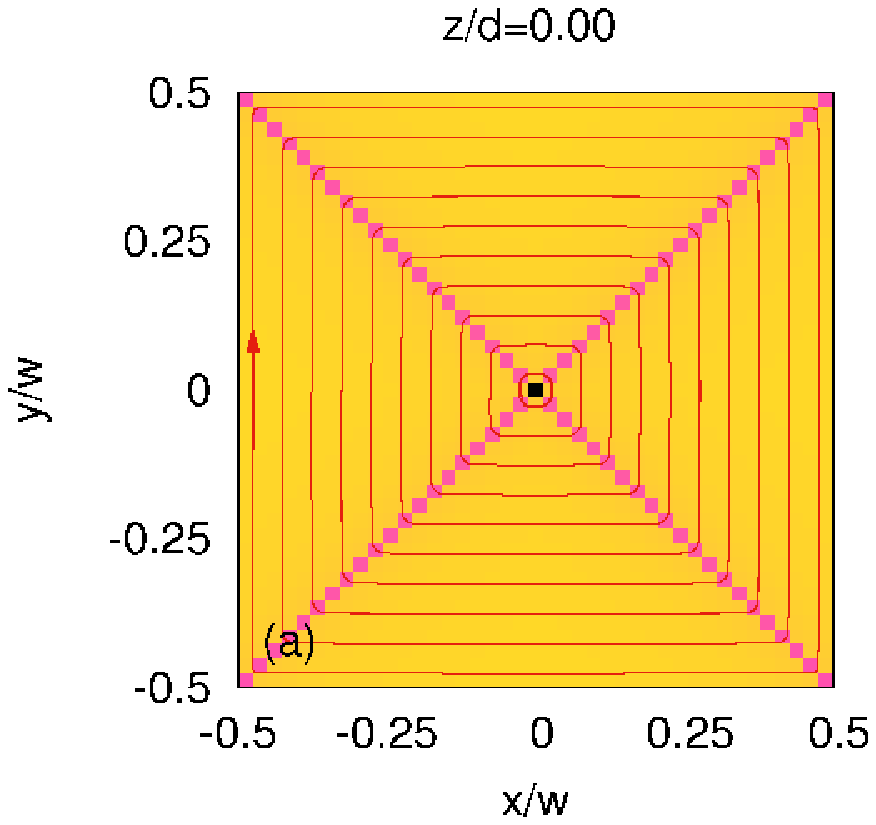}}
{\includegraphics[trim=0 -10 0 0,clip,height=4.6 cm]{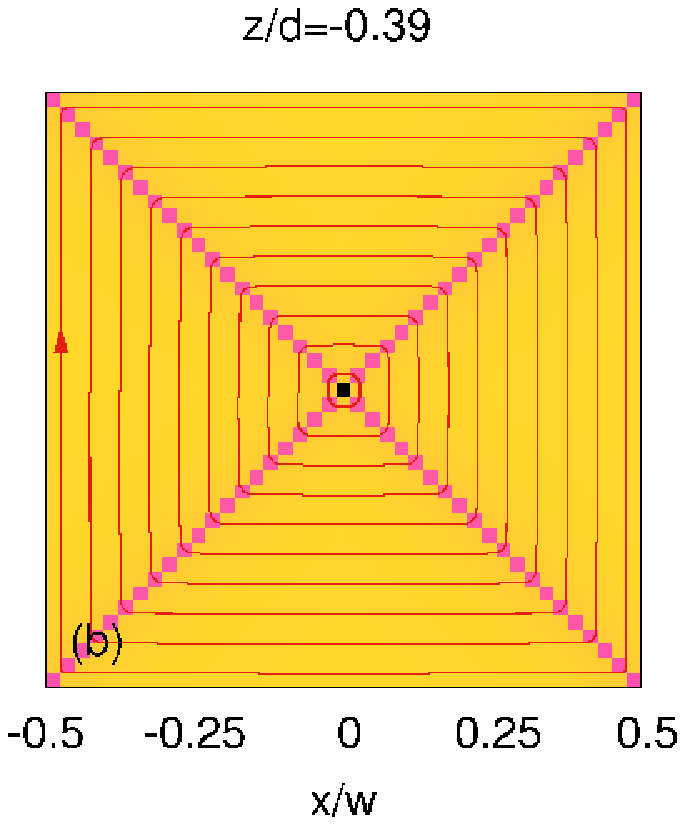}}  
{\includegraphics[trim=0 -10 0 0,clip,height=4.6 cm]{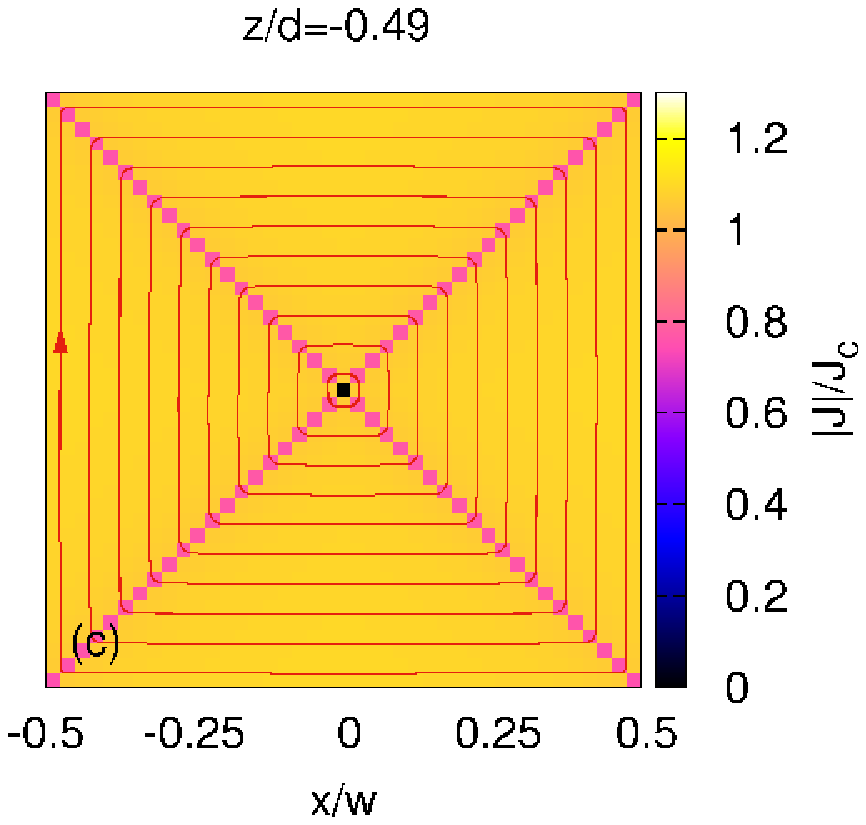}} \\
{\includegraphics[trim=0 0 0 0,clip,height=4.5 cm]{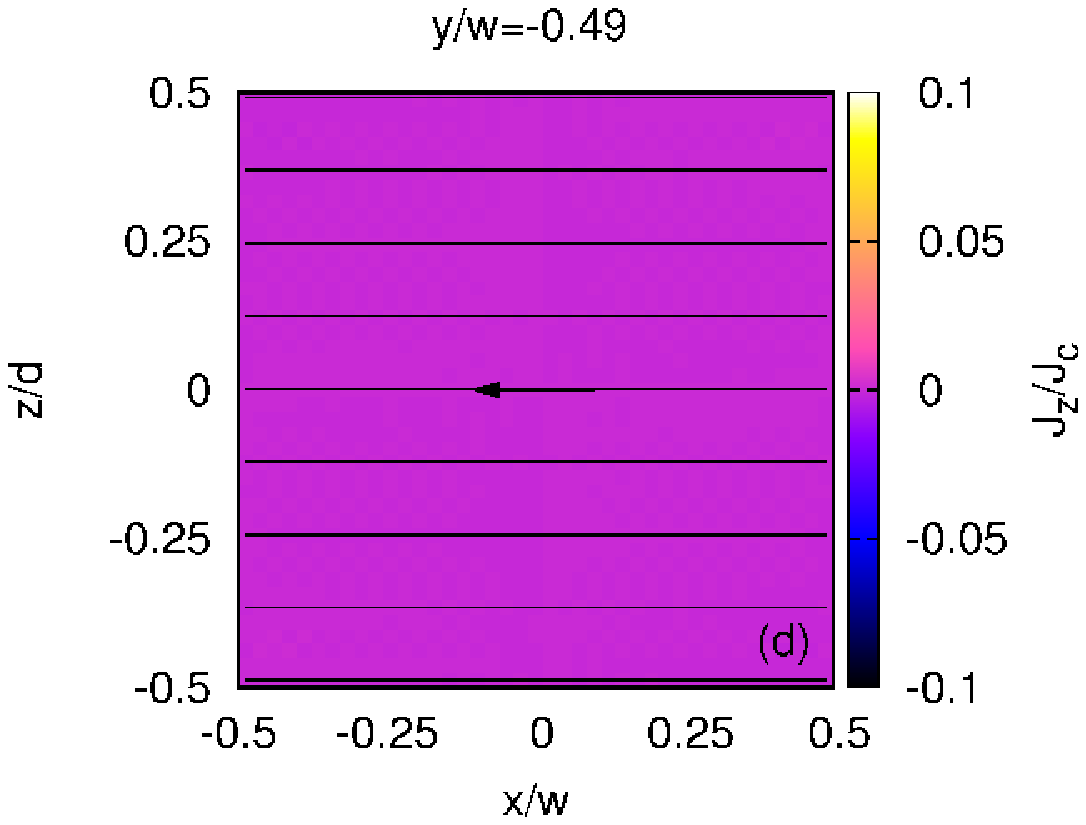}} 
{\includegraphics[trim=0 0 0 0,clip,height=4.5 cm]{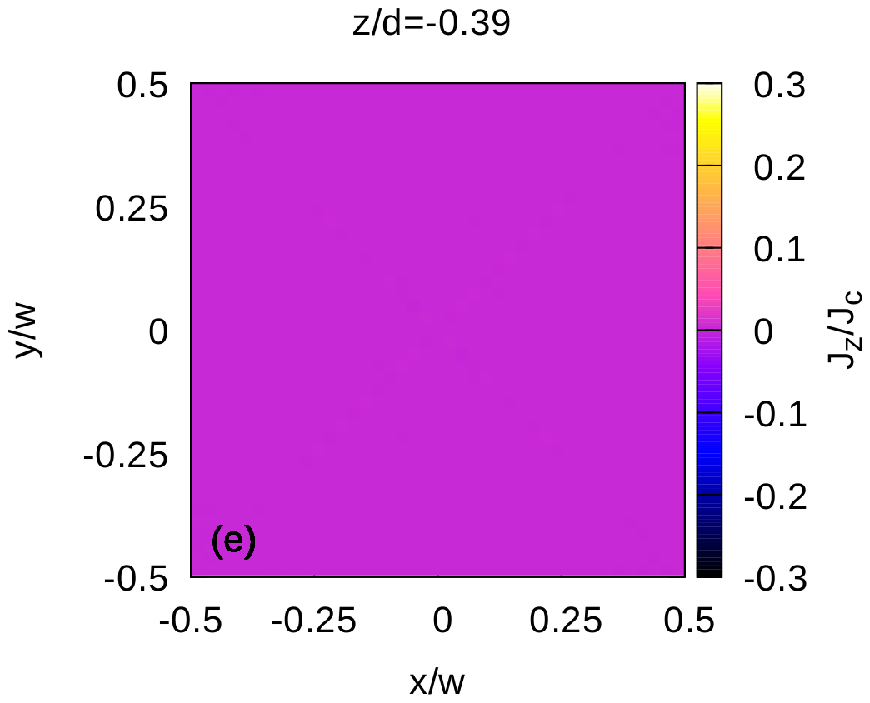}}
\caption{The same as figure \ref{cubeJc.fig} for (a,b,c,d) and figure \ref{cubecJz.fig}a for (e) but for an applied field $B_a=2.0B_s$. At high applied fields, the current lines are square and $J_z$ vanishes, as in CSM predictions for long bars.}\label{cubeJc2Bp.fig}
\end{figure}

For a cube, the current flux lines are practically square at the center (figure \ref{cubeJc.fig}a). Close to the ends, they change progressively from square next to the side surfaces to circular at the center (figure \ref{cubeJc.fig}c). At the intermediate height $z/d=-0.39$, there is a slight bending of the current lines (figure \ref{cubeJc.fig}b). The non-zero $J_z$ bends the current lines vertically, as seen in figure \ref{cubeJc.fig}d for a plane close to the side surface. After increasing the applied field to the effective saturation field, defined in section \ref{s.BpM}, the magnitude of $J_z$ reduces and the current flux lines become closer to squares (figure \ref{cubeJcBp.fig}). Well above the effective saturation field, the current flux lines are perfectly square (figure \ref{cubeJc2Bp.fig}).

\begin{figure}[ptb]
\centering
{\includegraphics[trim=0 -10 0 0,clip,height=4.6 cm]{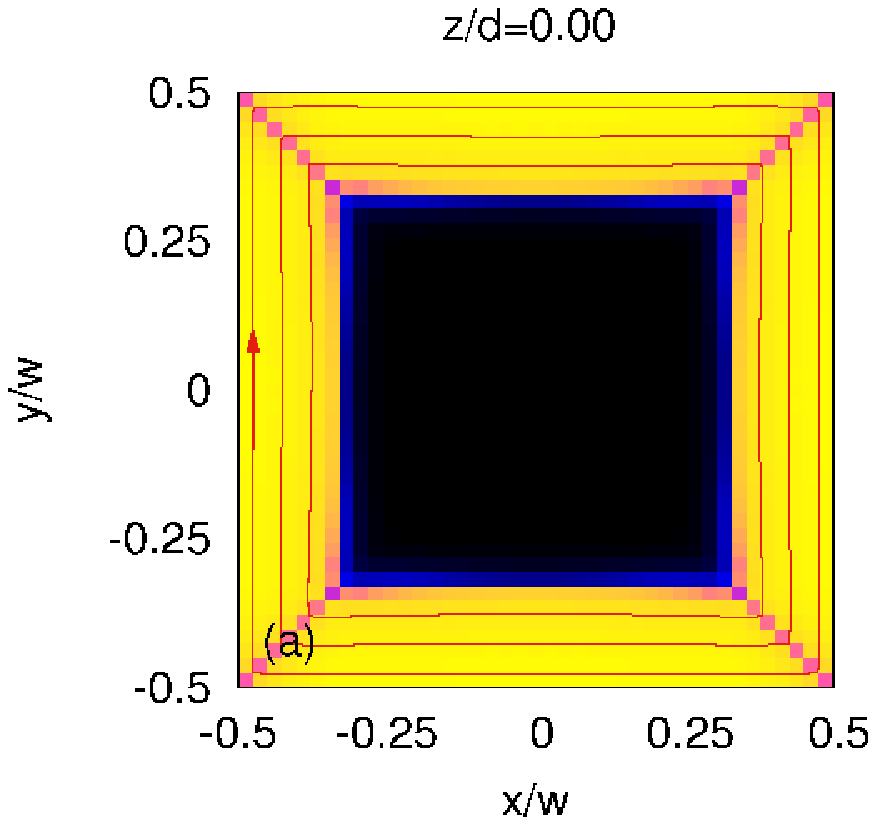}}
{\includegraphics[trim=0 -10 0 0,clip,height=4.6 cm]{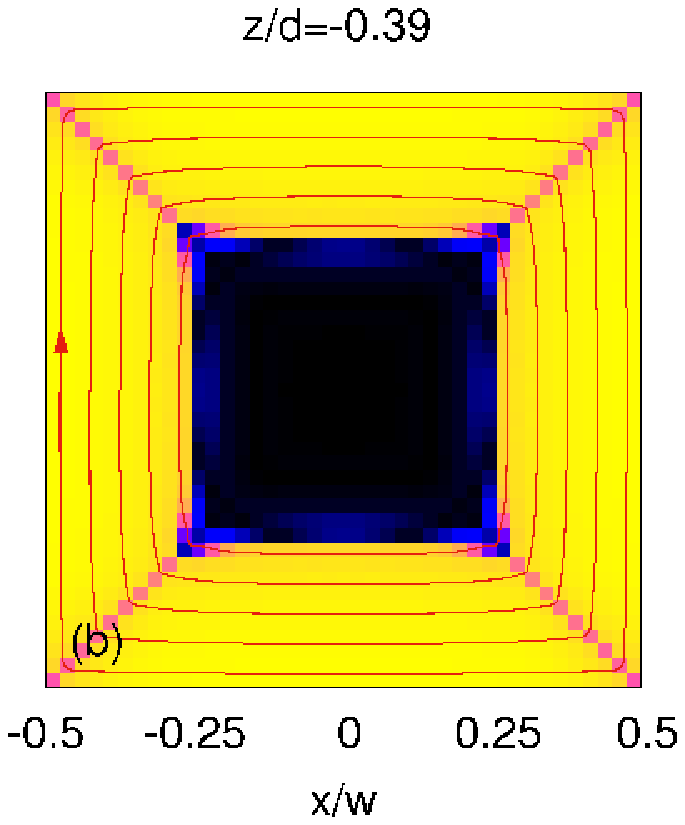}}  
{\includegraphics[trim=0 -10 0 0,clip,height=4.6 cm]{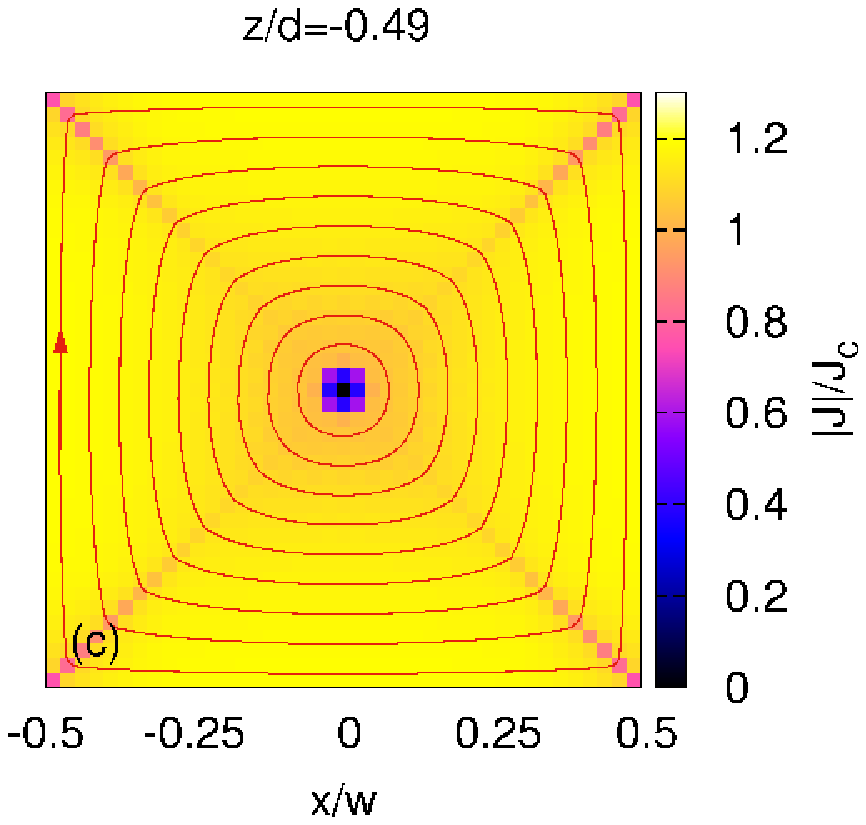}} \\
{\includegraphics[trim=0 0 0 0,clip,height=4.5 cm]{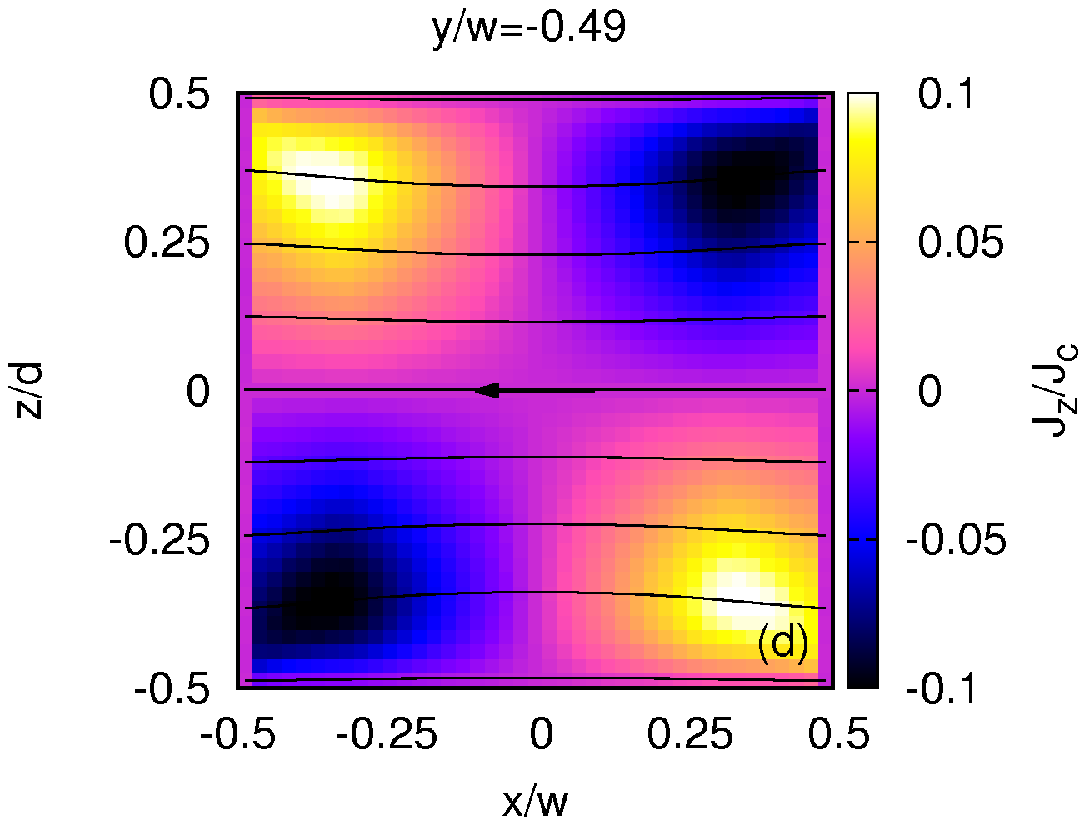}} 
{\includegraphics[trim=0 0 0 0,clip,height=4.5 cm]{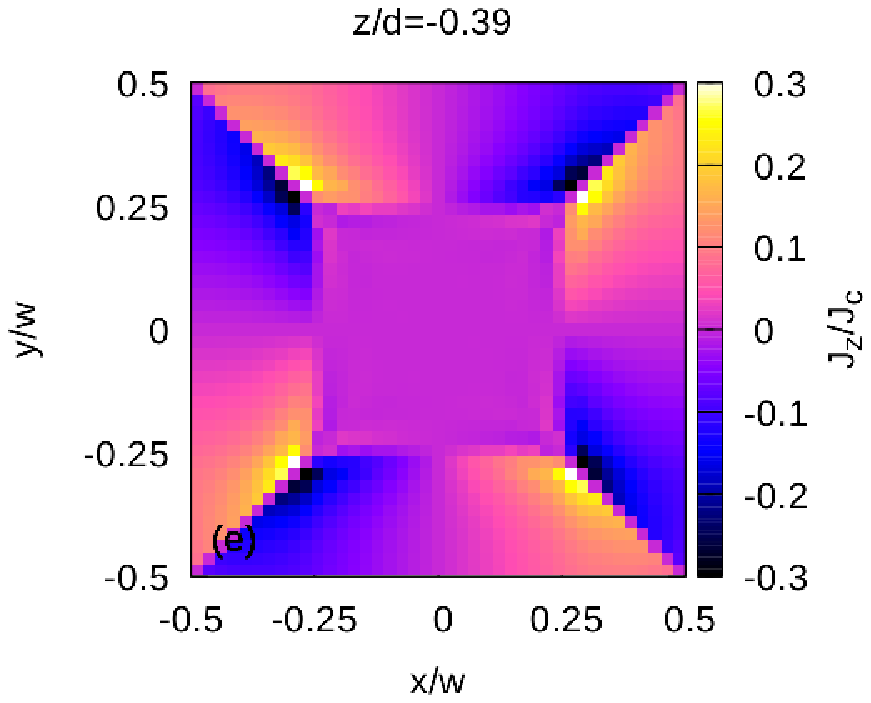}}
\caption{Same situation as figure \ref{cubeJc.fig} for (a,b,c,d) and figure \ref{cubecJz.fig}a for (e) but with power-law exponent 30.}\label{cubeJcn30.fig}
\end{figure}

\begin{figure}[ptb]
\centering
{\includegraphics[trim=0 0 0 0,clip,height=4.8 cm]{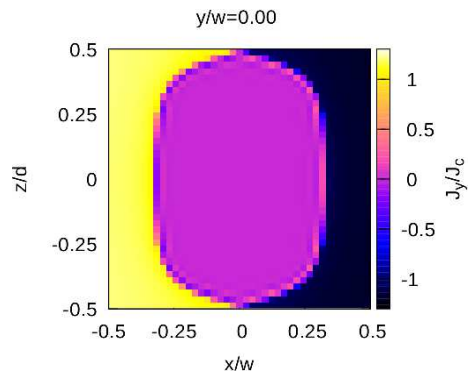}} 
\caption{The same as figure \ref{cubecJy.fig}a but for power-law exponent 30.}\label{cubeJcn30Jy.fig}
\end{figure}

For a cube with lower power-law exponent, $n=30$, the current density is approximately the same as for $n=100$. Then, the finite power-law exponent is not the cause of the $J_z$ component, since samples with substantially different power-law exponent present $J_z$ of similar magnitude (see figures \ref{cubecJz.fig}a and \ref{cubeJcn30.fig}e). On the contrary, the modulus of $\vJ$ for $n=30$ is larger that for $n=100$, overcoming substantially $J_c$ (figures \ref{cubeJc.fig}abc,\ref{cubeJcn30.fig}abc and figures \ref{cubecJy.fig}a,\ref{cubeJcn30Jy.fig}). The cause is that, for the same electric field caused by the applied field variation, lower $n$ exponents result in higher current densities. For the CSM, the maximum $|\vJ|$ will be $J_c$ and $|J_z|$ will be slighlty lower, specially away from the diagonals.


\subsection{Magnetization loops and saturation field}
\label{s.BpM}

\begin{figure}[ptb]
\centering
{\includegraphics[trim=0 0 0 0,clip,width=10.5 cm]{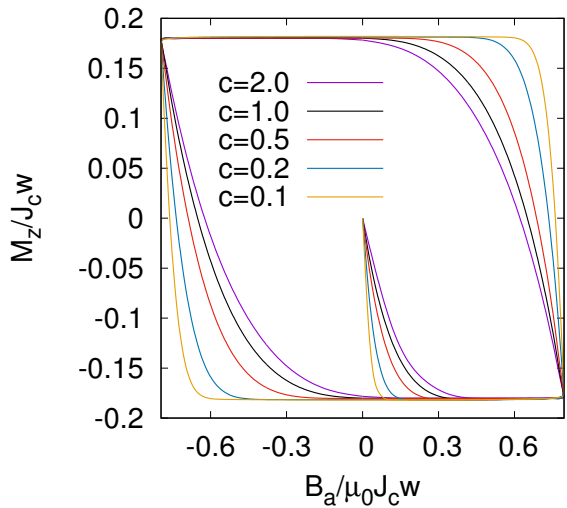}}
\caption{Magnetization loops of square bars with constant $J_c$ and several thickness-to-width aspect ratios $c=d/w$.}
\label{Magnetization.fig}
\end{figure}

We calculated the magnetization loops for several aspect ratios ${c\equiv d/w}$, assuming constant $J_c$. Figure \ref{Magnetization.fig} shows the hysteresis loops for $c=$1,0.5,0.2,0.1 and 1 T of peak applied field. From these loops, we obtained the saturation field, $B_s$, in table \ref{t.Bp}; which we define as the applied magnetic field that causes $M=0.99 M_s$, $M_s$ being the saturation magnetization. The value of $|M_s|$ is taken as the maximum $|M|$ at the initial curve. This definition of $B_s$ is more relevant than previous ones, being the applied field that fully saturates the sample with critical current \cite{bean62PRL} or the self-field at the sample center at full current penetration \cite{forkl94PhC}. The cause is that $M$ for thin samples practically saturates at applied fields much lower than $B_s$ from the previous definitions \cite{pardo04SST}. Indeed, a perfectly thin film requires an infinite applied field to be fully saturated with current \cite{halse70JPD,Brandt93PRBa,Zeldov94PRB}. The definition of $B_s$ as $M(B_s)=0.99M_s$ is practical for experimentalists, since it is tolerant to measurement errors in the $M(B_a)$ curve. Note that for our definition of saturation field, the sample is actually not fully penetrated by current (figure \ref{cubeJcBp.fig}), since the current loops close to the cube axis parallel to $z$ contribute very little to the magnetic moment. The finite power-law exponent of 100 slightly enhances this effect because $|\vJ|$ is slightly above $J_c$ at certain regions.

\begin{figure}[ptb]
\centering
{\includegraphics[trim=0 0 0 0,clip,width=10.5 cm]{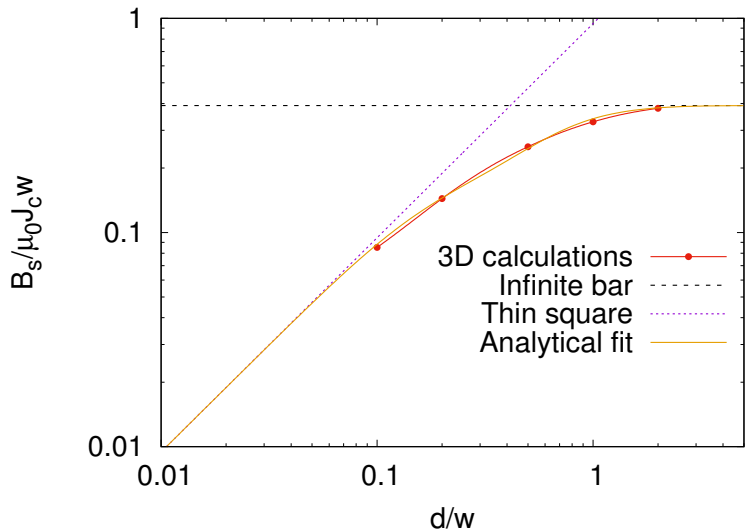}}
\caption{The computed saturation field, defined as $M(B_a)=0.99M_s$, being $M_s$ the saturation magnetization, approaches to the infinite bar and thin limits for high and low thickness-over-width ratio, $d/w$, respectively. The computations are done with power-law exponent 100; the infinite-bar limit is calculated from the magnetization curve in the CSM from \cite{chen89JAP}. The analytical fit of (\ref{Bpfit}) differs from the calculations by less than 3 \%.}
\label{Hp.fig}
\end{figure}

The saturation field increases with the aspect ratio (figure \ref{Hp.fig}) because, for the same width, thinner prisms generate lower maximum self-fields. Figure \ref{Hp.fig} also shows the infinite bar limit, calculated from the analytical magnetization curve of the CSM \cite{chen89JAP}, and the thin square situation, computed by our model. The saturation field of the latter is proportional to the film thickness. The results for finite thickness approach to the infinite bar and thin limits for high and low $c=d/w$ aspect ratios, respectively. The following analytical fit agrees exactly with both infinite and thin limits and differs from the intermediate cases by less than 3 \%
\begin{equation}
B_s(c)=\mu_0J_cw a_1 \left[1+a_2\exp\left\{ \frac{ -\ln^2(a_3c)}{2a_4^2} \right\}\right]\tanh(a_5c)
\label{Bpfit}
\end{equation}
with $a_1=0.3915$, $a_2=-0.26$, $a_3=2.56$, $a_4=0.75$, and $a_5=2.41$.

The calcuated saturation field with a finite power-law exponent is expected to slightly differ with that from the ideal CSM. Tests for 2D cross-sectional calculations for cylinders with the method in \cite{pardo15SST} show that the results with power-law exponent $n=100$ are over-estimated by only a few percent comparing to the CSM.

\begin{table}[tpb]
\begin{center}
\begin{tabular}{llll}
\hline
\hline
{\bf Aspect} & {\bf Saturation } \\
{\bf ratio ${d/w}$} & {\bf field ${B_{p}/{J_{c}}w\mu_{0}}$} \\
\hline
2.0 & 0.38 \\
1.0 & 0.32 \\
0.5 & 0.25 \\
0.2 & 0.14 \\
0.1 & 0.08 \\
\hline
\hline
\end{tabular}
\caption{The computed saturation field for square bars with finite thickness, defined as $M(B_s)=0.99M_s$; being $M_s$ the saturation magnetization. The limits for the thickness-to-width aspect ratio $d/w\gg 1$ and $d/w\ll 1$ are reproduced by equation (\ref{Bpfit}).}
\label{t.Bp}
\end{center}
\end{table}

\section{Conclusions}
\label{s.concl}

This article has presented a systematic study of the magnetization currents in rectangular prisms of several thickness-to-width aspect ratios. This study has been done by 3D electro-magnetic modeling by means of the variational method MEMEP3D \cite{memep3D}, which enables modeling the high number of degrees of freedom required for the computations.

For applied magnetic fields below the saturation field, we have found that the current lines do not follow rectangular paths. These current paths are possible thanks to bending in the direction of the applied magnetic field. Although the assumption of rectangular current paths may be fair to predict the magnetization loops, it will introduce significant errors in the prediction of generated magnetic field at the surface. Calculations for a relatively thin sample show that the average of the current density over the thickness is compatible with previous solutions for thin films and those extracted from magnetic field imaging experiments.

This article has also presented the magnetization loops and saturation magnetic field, defined by the applied field that causes a magnetization of 99 \% the saturation magnetization. This definition is more practical that previous ones, specially for thin prisms; previous definitions substantially over-estimating the saturation field.

The numerical computations of this article have provided an insight of current penetration. The 3D bending of the current lines indicates that there could be force-free and flux cutting effects \cite{clem11SST,badia15SST} in superconducting rectangular prisms. The component of the critical current density in the applied field direction may play a role, being the magnetization currents in a bulk and a stack of tapes not identical.

The possibility to solve 3D situations with a relatively high number of degrees of freedom enables to systematically analyze other situations, such as cross-field demagnetization or coupling effects in multi-filamentary tapes.


\section*{Acknowledgements}

The authors acknowledge the use of resources provided by the SIVVP project (ERDF, ITMS 26230120002) and the finantial support of the Grant Agency of the Ministry of Education of the Slovak Republic and the Slovak Academy of Sciences (VEGA) under contract no. 2/0126/15, as well as the R\&D Operational Program funded by the ERDF under Grant ITMS 26240120019 ``CENTE II"(0.5).


\appendix




\begin{thebibliography}{10}

\bibitem{hull04PIE}
J.~R. Hull and M.~Murakami.
\newblock Applications of bulk high-temperature superconductors.
\newblock {\em Proceedings of the IEEE}, 92(10):1705--1718, 2004.

\bibitem{murakami07ACT}
M.~Murakami.
\newblock Processing and applications of bulk {RE--Ba--Cu--O} superconductors.
\newblock {\em International journal of applied ceramic technology},
  4(3):225--241, 2007.

\bibitem{zhouD12SST}
D.~Zhou, M.~Izumi, M.~Miki, B.~Felder, T.~Ida, and M.~Kitano.
\newblock An overview of rotating machine systems with high-temperature bulk
  superconductors.
\newblock {\em Supercond. Sci. Technol.}, 25(10):103001, 2012.

\bibitem{werfel12SST}
F.~N. Werfel, U.~Floegel-Delor, R.~Rothfeld, T.~Riedel, B.~Goebel, D.~Wippich,
  and P.~Schirrmeister.
\newblock Superconductor bearings, flywheels and transportation.
\newblock {\em Supercond. Sci. Technol.}, 25(1):014007, 2012.

\bibitem{hecher15SST}
J.~Hecher, T.~Baumgartner, J.~D. Weiss, C.~Tarantini, A.~Yamamoto, J.~Jiang,
  E.~E. Hellstrom, D.~C. Larbalestier, and M.~Eisterer.
\newblock Small grains: a key to high-field applications of granular {Ba-122}
  superconductors?
\newblock {\em Supercond. Sci. Technol.}, 29(2):025004, 2015.

\bibitem{mishev15SST}
V~Mishev, M~Zehetmayer, DX~Fischer, M~Nakajima, H~Eisaki, and M~Eisterer.
\newblock Interaction of vortices in anisotropic superconductors with isotropic
  defects.
\newblock {\em Supercond. Sci. Technol.}, 28(10):102001, 2015.

\bibitem{tomita03Nat}
M.~Tomita and M.~Murakami.
\newblock High-temperature superconductor bulk magnets that can trap magnetic
  fields of over 17 tesla at 29 {K}.
\newblock {\em Nature}, 421(6922):517--520, 2003.

\bibitem{durrell14SST}
J.~H. Durrell, A.~R. Dennis, J.~Jaroszynski, M.~D. Ainslie, K.~G.~B. Palmer,
  Y.~H. Shi, A.~M. Campbell, J.~Hull, M.~Strasik, E.~E. Hellstrom, et~al.
\newblock A trapped field of 17.6 {T} in melt-processed, bulk {Gd-Ba-Cu-O}
  reinforced with shrink-fit steel.
\newblock {\em Supercond. Sci. Technol.}, 27(8):082001, 2014.

\bibitem{fuchs13SST}
G.~Fuchs, W.~H{\"a}{\ss}ler, K.~Nenkov, J.~Scheiter, O.~Perner, A.~Handstein,
  T.~Kanai, L.~Schultz, and B.~Holzapfel.
\newblock High trapped fields in bulk {MgB$_2$} prepared by hot-pressing of
  ball-milled precursor powder.
\newblock {\em Supercond. Sci. Technol.}, 26(12):122002, 2013.

\bibitem{ainslie15SST}
M.~D. Ainslie and H.~Fujishiro.
\newblock Modelling of bulk superconductor magnetization.
\newblock {\em Supercond. Sci. Technol.}, 28(5):053002, 2015.

\bibitem{acreview}
F.~Grilli, E.~Pardo, A.~Stenvall, D.~N. Nguyen, W.~Yuan, and F.~G{\"om\"o}ry.
\newblock Computation of losses in {HTS} under the action of varying magnetic
  fields and currents.
\newblock {\em IEEE Trans. Appl. Supercond.}, 24(1):8200433, 2014.

\bibitem{chen89JAP}
D.-X. Chen and R.~B. Goldfarb.
\newblock Kim model for magnetization of type-{II} superconductors.
\newblock {\em J. Appl. Phys.}, 66(6):2489--2500, 1989.

\bibitem{brandt95PRL}
E.H. Brandt.
\newblock Square and rectangular thin superconductors in a transverse magnetic
  field.
\newblock {\em Phys. Rev. Lett.}, 74(15):3025--3028, 1995.

\bibitem{brandt95PRBa}
E.H. Brandt.
\newblock Electric field in superconductors with rectangular cross section.
\newblock {\em Phys. Rev. B}, 52(21):15442, 1995.

\bibitem{prigozhin98JCP}
L.~Prigozhin.
\newblock Solution of thin film magnetization problems in {type-II}
  superconductivity.
\newblock {\em J. Comput. Phys.}, 144(1):180--193, 1998.

\bibitem{navau08JAP}
C.~Navau, A.~Sanchez, N.~Del-Valle, and D.~X. Chen.
\newblock Alternating current susceptibility calculations for thin-film
  superconductors with regions of different critical-current densities.
\newblock {\em J. Appl. Phys.}, 103:113907, 2008.

\bibitem{pecher04ICS}
R.~Pecher, M.D. McCulloch, S.J. Chapman, L.~Prigozhin, and C.M. Elliott.
\newblock {3D}-modelling of bulk type{-II} superconductors using unconstrained
  {H-}formulation.
\newblock {\em Inst. of Phys.: Conf. Ser.}, 181:1418, 2003.
\newblock European Conference on Applied Superconductivity ({EUCAS}) 2003.

\bibitem{badia05APL}
A.~Bad{\'\i}a-Maj{\'o}s and C.~L{\'o}pez.
\newblock Critical state model in superconducting parallelepipeds.
\newblock {\em Appl. Phys. Lett.}, 86(20):202510, 2005.

\bibitem{zhangM12SSTa}
M.~Zhang and TA~Coombs.
\newblock {3D} modeling of high{-$T_c$} superconductors by finite element
  software.
\newblock {\em Supercond. Sci. Technol.}, 25:015009, 2012.

\bibitem{lousberg09SST}
G.P. Lousberg, M.~Ausloos, C.~Geuzaine, P.~Dular, P.~Vanderbemden, and
  B.~Vanderheyden.
\newblock Numerical simulation of the magnetization of high-temperature
  superconductors: a {3D} finite element method using a single time-step
  iteration.
\newblock {\em Supercond. Sci. Technol.}, 22:055005, 2009.

\bibitem{campbell14SSTa}
A.~M. Campbell.
\newblock Solving the critical state using flux line properties.
\newblock {\em Supercond. Sci. Technol.}, 27(12):124006, 2014.

\bibitem{fagnard16SST}
J.-F. Fagnard, M.~Morita, S.~Nariki, H.~Teshima, H.~Caps, B.~Vanderheyden, and
  P.~Vanderbemden.
\newblock Magnetic moment and local magnetic induction of
  superconducting/ferromagnetic structures subjected to crossed fields:
  experiments on {GdBCO} and modelling.
\newblock {\em Supercond. Sci. Technol.}, 29(12):125004, 2016.

\bibitem{komi09PhC}
Y.~Komi, M.~Sekino, and H.~Ohsaki.
\newblock Three-dimensional numerical analysis of magnetic and thermal fields
  during pulsed field magnetization of bulk superconductors with inhomogeneous
  superconducting properties.
\newblock {\em Physica C}, 469(15):1262--1265, 2009.

\bibitem{ainslie14SST}
M.~D. Ainslie, H.~Fujishiro, T.~Ujiie, J.~Zou, A.~R. Dennis, Y.~H. Shi, and
  D.~A. Cardwell.
\newblock Modelling and comparison of trapped fields in (re) bco bulk
  superconductors for activation using pulsed field magnetization.
\newblock {\em Supercond. Sci. Technol.}, 27(6):065008, 2014.

\bibitem{grilli05IES}
F.~Grilli, S.~Stavrev, Y.~Le~Floch, M.~Costa-Bouzo, E.~Vinot, I.~Klutsch,
  G.~Meunier, P.~Tixador, and B.~Dutoit.
\newblock Finite-element method modeling of superconductors: from {2-D} to
  {3-D}.
\newblock {\em IEEE Trans. Appl. Supercond.}, 15(1):17--25, 2005.

\bibitem{memep3D}
E~Pardo and M~Kapolka.
\newblock {3D} computation of non-linear eddy currents: variational method and
  superconducting cubic bulk.
\newblock arXiv:1611.04752 [cond-mat.supr-con].

\bibitem{jooss02RPP}
C.~Jooss, J.~Albrecht, H.~Kuhn, S.~Leonhardt, and H.~Kronmuller.
\newblock Magneto-optical studies of current distributions in {high-$T_c$}
  superconductors.
\newblock {\em Rep. Prog. Phys.}, 65:651--788, 2002.

\bibitem{romerosalazar10PRB}
C.~Romero-Salazar, C.~Jooss, and O.~A. Hernandez-Flores.
\newblock Reconstruction of the electric field in {type-II} superconducting
  thin films in perpendicular geometry.
\newblock {\em Phys. Rev. B}, 81(14):144506, 2010.

\bibitem{wells17ScR}
F.~S. Wells, A.~V. Pan, I.~A. Golovchanskiy, S.~A. Fedoseev, and A.~Rozenfeld.
\newblock Observation of transient overcritical currents in {YBCO} thin films
  using high-speed magneto-optical imaging and dynamic current mapping.
\newblock {\em Scientific Reports}, 7:40235, 2017.

\bibitem{brandt98PRBa}
E.~H. Brandt.
\newblock Superconductor disks and cylinders in an axial magnetic field. {I}.
  {Flux} penetration and magnetization curves.
\newblock {\em Phys. Rev. B}, 58(10):6506, 1998.

\bibitem{sanchez01PRB}
A.~Sanchez and C.~Navau.
\newblock Magnetic properties of finite superconducting cylinders. {I.} uniform
  applied field.
\newblock {\em Phys. Rev. B}, 64:214506, 2001.

\bibitem{bean62PRL}
C.~P. Bean.
\newblock Magnetization of hard superconductors.
\newblock {\em Phys. Rev. Lett.}, 8(6):250--253, 1962.

\bibitem{forkl94PhC}
A.~Forkl and H.~Kronm{\"u}ller.
\newblock A contribution to the analysis of the current-density distribution in
  elongated hard type{-II} superconductors with rectangular cross-section.
\newblock {\em Physica C}, 228(1):1--14, 1994.

\bibitem{pardo04SST}
E.~Pardo, D.-X. Chen, A.~Sanchez, and C.~Navau.
\newblock The transverse critical-state susceptibility of rectangular bars.
\newblock {\em Supercond. Sci. Technol.}, 17:537, 2004.

\bibitem{halse70JPD}
M.~R. Halse.
\newblock {AC} face field losses in a type {II} superconductor.
\newblock {\em J. Phys. D: Appl. Phys.}, 3:717--720, 1970.

\bibitem{Brandt93PRBa}
{E. H.} Brandt and M.~Indenbom.
\newblock {Type-II-superconductor} strip with current in a perpendicular
  magnetic field.
\newblock {\em Phys. Rev. B}, 48(17):12893--12906, 1993.

\bibitem{Zeldov94PRB}
E.~Zeldov, J.~R. Clem, M.~{McElfresh}, and M.~Darwin.
\newblock Magnetization and transport currents in thin superconducting films.
\newblock {\em Phys. Rev. B}, 49(14):9802--9822, 1994.

\bibitem{pardo15SST}
E.~Pardo, J.~{\v Souc}, and L.~{Frolek}.
\newblock Electromagnetic modelling of superconductors with a smooth
  current-voltage relation: variational principle and coils from a few turns to
  large magnets.
\newblock {\em Supercond. Sci. Technol.}, 28:044003, 2015.

\bibitem{clem11SST}
J.R. Clem, M.~Weigand, J.~H. Durrell, and A.~M. Campbell.
\newblock Theory and experiment testing flux-line cutting physics.
\newblock {\em Supercond. Sci. Technol.}, 24:062002, 2011.

\bibitem{badia15SST}
A.~Bad{\'\i}a-Maj{\'o}s and C.~L{\'o}pez.
\newblock Modelling current voltage characteristics of practical
  superconductors.
\newblock {\em Supercond. Sci. Technol.}, 28(2):024003, 2015.

\end{thebibliography}

\end{document}